\documentclass[a4paper,11pt]{article} 

\usepackage[margin=25mm]{geometry} 
\usepackage[numbers,sort&compress]{natbib}
\usepackage{hyperref} 

\usepackage{bm}
\usepackage{amsmath}
\usepackage{subcaption} 
\usepackage{ulem}

\usepackage{amsmath,amssymb,amsthm,amsfonts} 

\usepackage{graphicx} 
\usepackage{array}    
\usepackage{booktabs} 

\providecommand{\tsc}[1]{\textsc{\lowercase{#1}}}

\usepackage{tikz}
\usetikzlibrary{positioning, shapes.geometric, arrows.meta}
\usetikzlibrary{decorations.pathmorphing}

\newtheorem{remark}{Remark}
\newcommand{\te}{\mathsf{e}} 
\newcommand{\tf}{\mathsf{f}} 
\newcommand{\re}{\mathsf{E}} 
\newcommand{\rf}{\mathsf{F}} 
\newcommand{\rg}{\mathsf{G}} 
\newcommand{\de}{\mathfrak{e}} 
\newcommand{\df}{\mathfrak{f}} 
\newcommand{\dg}{\mathfrak{g}} 
\newcommand{\eref}{Eq.~\eqref}
\newcommand{\erefs}{Eqs.~\eqref}
\newcommand{\fref}{Fig.~\ref}
\newcommand{\frefs}{Figs.~\ref}
\newcommand{\tref}{Table~\ref}

\newcommand{\sref}{Section~\ref}
\newcommand{\saref}{Appendix~\ref}

\tikzset{
  StartEnd/.style = {rectangle, rounded corners, draw, minimum height=6mm, minimum width=20mm, text centered, fill=white},
  Process/.style  = {rectangle, draw, minimum height=7.5mm, text width=45mm, text centered, fill=white},
  Decision/.style = {diamond, draw, aspect=2.2, minimum height=8mm, inner sep=0pt, text centered, fill=white},
  Arrow/.style    = {thick, -{Stealth[length=2mm, width=2mm]}}
}

\def\tsc#1{\csdef{#1}{\textsc{\lowercase{#1}}\xspace}}
\tsc{WGM}
\tsc{QE}
\tsc{EP}
\tsc{PMS}
\tsc{BEC}
\tsc{DE}

\makeatletter
\def\ps@pprintTitle{%
    \let\@oddhead\@empty
    \let\@evenhead\@empty
    \def\@oddfoot{\reset@font\hfil\thepage\hfil}
    \let\@evenfoot\@oddfoot
}
\makeatother

\begin{document}
\let\WriteBookmarks\relax
\def\floatpagepagefraction{1}
\def\textpagefraction{.001}

\title{Topology optimization of conduction-radiation problems based on a ray-tracing approach}

\author{
  Shun Noguchi$^{1,*}$,
  Naoyuki Ishida$^{1}$,
  Jike Han$^{2,1}$,
  Kazuhiro Izui$^{1}$,
  Shinji Nishiwaki$^{3}$ \\
  \small $^1$Department of Micro Engineering, Kyoto University, Kyoto 615-8540, Japan \\
  \small $^2$The Hakubi Center for Advanced Research, Kyoto University, Kyoto 606-8501, Japan \\
  \small $^3$Department of Mechanical Engineering and Science, Kyoto University, Kyoto 615-8540, Japan \\
  \small $^*$Corresponding author: noguchi.shun.f36@kyoto-u.jp
}

\date{}
\maketitle

\begin{abstract}
  Thermal management is essential in space systems, where electronic devices must dissipate heat via radiative heat transfer.
  To achieve efficient designs of radiative cooling devices, structural optimization approaches such as topology optimization are required.
  While existing topology optimization methods have incorporated radiative heat transfer with certain simplifications, fully accounting for multidirectional mutual radiation remains challenging.
  To address this issue, this study proposes a density-based topology optimization method for conduction-radiation heat transfer problems that accounts for multidirectional mutual radiation.
  The proposed method integrates a zonal-method-based radiative heat transfer analysis incorporating a ray-tracing method into the finite element heat conduction analysis, capturing radiation effects during the optimization process.
  By treating the intermediate material densities that arise during the optimization as participating media, the proposed method enables a physically consistent evaluation of radiative heat transfer on implicitly represented structural boundaries.
  The analytical design sensitivities are derived using the adjoint method, and the accuracy is confirmed by the comparison with the numerical sensitivities obtained by the finite difference method.
  Numerical examples demonstrate the optimization of radiative heat sinks and radiation shields.
  The heat sink examples clarify how the balance between conduction and radiation governs the resulting designs, while the radiation shield examples produce multilayer insulation structures that are not obtained by conventional approaches.
\end{abstract}




\section{Introduction}\label{secIntroduction}
Thermal management is an essential technology in modern engineering systems.
Fundamentally, heat transfer is classified into three modes: (i) heat conduction, (ii) convection, and (iii) thermal radiation.
Heat conduction transfers thermal energy through materials via molecular interactions, convection transports heat through fluid motion, and thermal radiation exchanges thermal energy through electromagnetic waves.
In ordinary terrestrial environments, conduction and convection usually dominate thermal transport because surrounding fluids such as air and water efficiently remove heat from structures.
However, in the vacuum of space, convective heat transfer is unavailable due to the absence of a surrounding fluid medium.
Consequently, thermal radiation becomes the dominant mechanism for dissipating heat from electronic devices and structural components operating in space environments.
\par
\begin{figure}[htbp]
  \centering
  \includegraphics[scale=1.0]{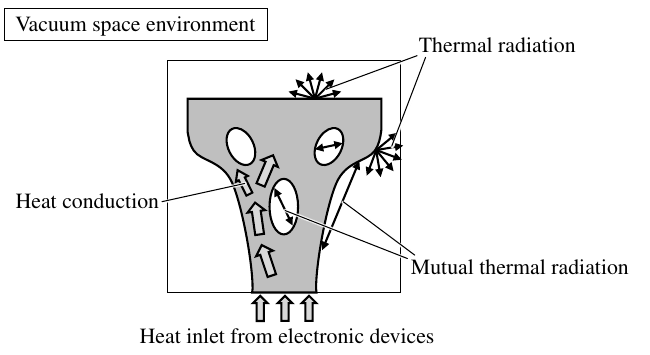}
  \caption{Concept of a radiative heat sink in a vacuum space environment.}
  \label{figConcept}
\end{figure}

For this reason, space systems are commonly equipped with radiative heat sinks that dissipate internally generated heat into outer space.
As illustrated in \fref{figConcept}, these devices transport the heat generated by electronic components through internal heat conduction and subsequently emit it into the vacuum environment via thermal radiation. Since launch mass and payload space are strictly limited in aerospace applications, such radiative cooling devices are required to be highly compact and lightweight while maintaining sufficient thermal performance. Therefore, achieving efficient thermal management under severe mass and volume constraints is crucial for ensuring stable device operation and extending device lifespan.
\par
Conventional designs of radiative cooling devices have typically relied on size and shape optimization techniques. In these approaches, the dimensions and arrangements of predefined structures, such as radiating fins and radiator panels, are parametrically optimized to maximize heat dissipation \cite{daun2003geometric, howell2003use, tan2011geometric, farahmand2012geometric, jang2012multidisciplinary}.
Although radiative heat transfer effects have been extensively incorporated into these conventional design frameworks, the achievable thermal performance remains limited by the initial geometric assumptions.
Therefore, to realize more thermally efficient structures, advanced design methodologies capable of generating high-performance configurations without relying on predefined geometries are required.
\par
Topology optimization (TO) is one of the most flexible structural optimization methods.
By representing the structure as a spatial distribution of materials, this approach can generate optimal designs while enabling the formation of holes.
TO was originally developed by Bends{\o}e \& Kikuchi \cite{bendsoeGeneratingOptimalTopologies1988} and has been extended to various physics problems \cite{silva1999design,borrvallTopologyOptimizationFluids2003,kiziltas_topology_2003,duhring2008acoustic}.
In the context of thermal management, TO has been developed for each of the three fundamental modes of heat transfer:
(i) heat conduction \cite{alberto2004topology,gersborg2006topologythermal, zhang2008design,gao2008topology,sun2026multi},
(ii) convection involving natural convection \cite{alexandersenTopologyOptimisationNatural2014, coffinLevelsetMethodSteadystate2016, alexandersenLargeScaleThreedimensional2016, fepponTopologyOptimizationThermal2020, liOptimumDesignThermal2022}
and forced convection \cite{dede2009multiphysics, yoonTopologicalDesignHeat2010, matsumoriTopologyOptimizationFluid2013, kogaDevelopmentHeatSink2013, yajiTopologyOptimizationMethod2015, yajiTopologyOptimizationThermalfluid2016, ghasemiMultiobjectiveTopologyOptimization2021, noguchi2026topology}, and
(iii) thermal radiation \cite{castro2015design}.
However, radiative cooling devices fundamentally rely on the coupled interaction between (i) internal heat conduction and (iii) external thermal radiation.
Therefore, designing such systems requires a TO framework that simultaneously considers both conduction and radiation (i.e., conduction-radiation problems), rather than treating each heat transfer mode independently.
In developing a TO framework for conduction-radiation problems, the density-based approach is particularly suitable. Since density-based methods are the most widely adopted approaches in TO, many existing studies on heat transfer optimization have been developed within this framework. Moreover, their implicit material representation provides high compatibility with finite element implementations and facilitates integration with other optimization problems. Given this versatility and extensibility, establishing a density-based TO framework for conduction-radiation problems is a promising research direction.
\par
\begin{table}[htbp]
  \centering
  \footnotesize
  \caption{Comparison of thermal radiation treatments and analysis methods integrated into density-based TO.}
  \label{tab:MethodComparison}
  \begin{tabular}{l|ccc}
    \toprule
    Approach                                                                                  & (1) Intermediate density & (2) Exact sensitivity        & (3) Computational efficiency \\
    \midrule
    Boundary conditions (used in \cite{bruns2007topology,cohen2022level,onodera2025topology}) & $\times$                 & $\bigcirc$                   & $\bigcirc$                   \\
    Net-radiation method (used in \cite{sevart2022dual})                                      & $\times$                 & $\bigcirc$                   & $\bigcirc$                   \\
    Monte Carlo ray tracing  (used in \cite{sas2023design})                                   & $\bigcirc$               & $\times$ (Statistical noise) & $\times$                     \\
    Discrete ordinates / Finite volume                                                        & $\bigcirc$               & $\bigcirc$                   & $\times$ (High-dimensional)  \\
    \midrule
    \textbf{Zonal method (used in Present work)}                                              & $\bigcirc$               & $\bigcirc$                   & $\bigcirc$                   \\
    \bottomrule
  \end{tabular}
\end{table}
However, integrating thermal radiation into a density-based TO framework presents significant challenges, primarily due to the characteristics of radiative heat transfer analysis methods.
Existing approaches for conduction-radiation analysis exhibit different trade-offs in terms of accuracy, computational cost, and compatibility with density-based TO, as summarized in \tref{tab:MethodComparison}.
One approach is to treat thermal radiation simply as boundary conditions in heat conduction analysis, where radiative heat transfer is represented as heat fluxes on structural surfaces \cite{bruns2007topology,cohen2022level,onodera2025topology}. Although computationally efficient, this simplification fails to evaluate multidirectional mutual radiation, as it neglects the geometric relationships among the surfaces of neighboring objects.
The net-radiation method calculates radiative heat transfer more accurately using view factors, which describe the geometric relationships among surfaces \cite{sparrow2018radiation,howell2020thermal}. While capable of evaluating mutual radiation, this method requires explicit surface geometries, making it difficult to adapt to the implicit surface representation in density-based TO.
To address this difficulty, Sevart \& Bergman \cite{sevart2022dual} proposed a TO approach called the dual method, which incorporates a binarization process to identify radiative surfaces during the optimization process. However, this non-differentiable binarization process prohibits the exact calculation of design sensitivities regarding radiative heat transfer.
The Monte Carlo ray-tracing method is a highly accurate statistical approach that generates and tracks numerous rays \cite{howell1998monte,modest2003backward}. However, its inherent statistical noise poses a severe challenge for exact sensitivity calculations. While Sas Brunser \& Steinfeld \cite{sas2023design} successfully utilized this method within a gradient-free TO framework, such non-gradient approaches restrict the scale of computationally tractable problems.
Alternatively, partial differential equation-based approaches, such as the discrete ordinates method and the finite volume method, solve the radiative transfer equation by discretizing the angular space \cite{stamnes1981new,evans1998spherical,coelho2014advances}. Although these methods are compatible with grid-based analyses and thereby density-based TO approaches, they require iteratively solving a high-dimensional problem involving three spatial and two angular dimensions at each optimization step, imposing a prohibitively high computational burden.
\par
As indicated by these prior studies, TO for conduction-radiation problems continues to face the difficulties regarding the evaluation of multidirectional mutual radiation \cite{bruns2007topology,cohen2022level,onodera2025topology}, the derivation of exact design sensitivities \cite{sevart2022dual}, and the management of computational costs \cite{sas2023design}.
To overcome these challenges simultaneously, a radiative heat transfer analysis method must be properly integrated into the density-based TO framework while satisfying the following requirements:
\begin{enumerate}
  \item Since intermediate material densities between solid and void may emerge in density-based TO, the analysis method must be capable of evaluating multidirectional mutual radiation even in intermediate-density regions.
  \item Since gradient-based optimization is employed, the analysis method must avoid non-differentiable formulations in the forward analysis to enable the derivation of design sensitivities.
  \item Since the forward analysis is repeatedly solved throughout the optimization process, the analysis method must maintain a sufficiently low computational cost while allowing sufficiently high design resolution.
\end{enumerate}
Among various radiative heat transfer analysis approaches in \tref{tab:MethodComparison}, the zonal method \cite{hottel1958radiant,modest2021radiative} has the potential to satisfy these requirements.
This method evaluates thermal radiation by dividing the computational domain into discrete volume zones and calculating the radiative heat transfer between them via exchange factors.
Unlike partial differential equation-based angular discretization methods \cite{stamnes1981new,evans1998spherical,coelho2014advances}, the zonal method evaluates radiation exchange through precomputed exchange factors between zones, avoiding the iterative solution of high-dimensional angular fields.
Crucially, this approach accounts for participating media, namely media that absorb, emit, and scatter thermal radiation within the spatial domain.
This characteristic makes the zonal method potentially compatible with density-based TO. To realize such an analysis within density-based TO, however, a novel method to evaluate the exchange factors with respect to the intermediate densities that emerge during the optimization is required.
\par
Against this background, this study proposes a density-based TO method for conduction-radiation problems, in which the exchange factors of the zonal method are evaluated using differentiable ray tracing.
To incorporate this procedure into the density-based framework, this study introduces two key technical contributions:
\begin{enumerate}
  \item
        A participating medium model is proposed for continuously interpolating solid and void regions, thereby enabling the proper evaluation of multidirectional mutual radiation even in intermediate-density regions.
  \item
        A differentiable ray-tracing formulation is developed for evaluating exchange factors within spatially varying density fields, enabling the derivation of analytically consistent design sensitivities for gradient-based TO.
\end{enumerate}
Consequently, the proposed ray-tracing-based formulation enables a differentiable and physically consistent evaluation of multidirectional mutual radiation within the density-based TO framework, providing an effective approach to the thermal management demands of space applications.
\par
The remainder of this paper is organized as follows.
\sref{secFormulation} describes the formulation of the conduction-radiation heat transfer analysis and the proposed methodology.
\sref{secNumericalExample} demonstrates the effectiveness of the proposed method through a series of numerical examples, including the design of radiative heat sinks and radiation shields.
Finally, \sref{secConclusion} concludes the paper.

\section{Formulation}\label{secFormulation}
This section presents the mathematical formulation and numerical implementation for the conduction-radiation TO.
\sref{secDesignVariables} describes the spatial discretization scheme and the definition of design variables.
\sref{secGoverningEquation} formulates the governing equations for the conduction-radiation problem.
Subsequently, \sref{secExchangeFactor} derives the exchange factors required to evaluate radiative heat transfer.
\sref{secDiscretization} presents the discretization of the governing equations for numerical implementation.
Finally, \sref{secTopologyOptimization} formulates the TO problem and the corresponding adjoint sensitivity analysis.
\begin{remark}
  To avoid ambiguity in the subsequent formulation, different index styles are adopted according to their roles.
  Italic lowercase indices (e.g., $a$, $b$, $c$, $\ldots$) denote generic indices used in algebraic and tensor operations.
  Sans-serif lowercase indices (e.g., $\te$, $\mathsf{f}$, $\ldots$) are used to identify reference finite elements under consideration.
  Sans-serif uppercase indices (e.g., $\re$, $\rf$, $\ldots$) denote finite elements interacting with the corresponding reference element, particularly in the evaluation of radiative heat transfer.
  In addition, Fraktur lowercase indices (e.g., $\de$, $\df$, $\ldots$) denote the finite elements with respect to which design sensitivities are calculated.
\end{remark}
\subsection{Design variables}\label{secDesignVariables}
\begin{figure}[htbp]
  \centering
  \includegraphics[scale=1.0]{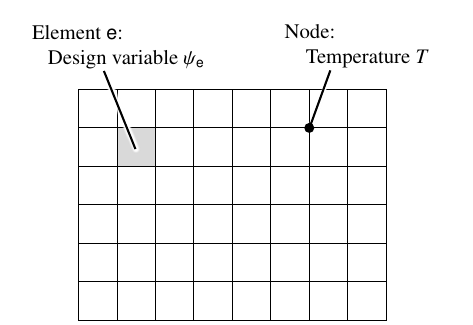}
  \caption{Concept of spatial discretization of analysis domain.}
  \label{figDiscretization}
\end{figure}
As illustrated in \fref{figDiscretization}, the analysis domain $\Omega$ is discretized using a structured mesh.
Let $N_{\rm elem}$ and $N_{\rm node}$ denote the number of elements and nodes in the discretized domain, respectively.
A design variable $\psi_{\tf}$ is assigned to each element $\tf \,(\tf = 1, 2, \dots, N_{\rm elem})$.
These design variables are expressed in vector form as $\bm{\psi} = \{\psi_1, \psi_2, \dots, \psi_{N_{\rm elem}}\}^\top$.
Each design variable is bounded between $0$ and $1$, as follows:
\begin{align}\label{eqdesignVariableBounds}
  0 \le \psi_{\tf} \le 1 \quad (\tf = 1, 2, \dots, N_{\rm elem}).
\end{align}
\begin{figure}[htbp]
  \centering
  \includegraphics[scale=1.0]{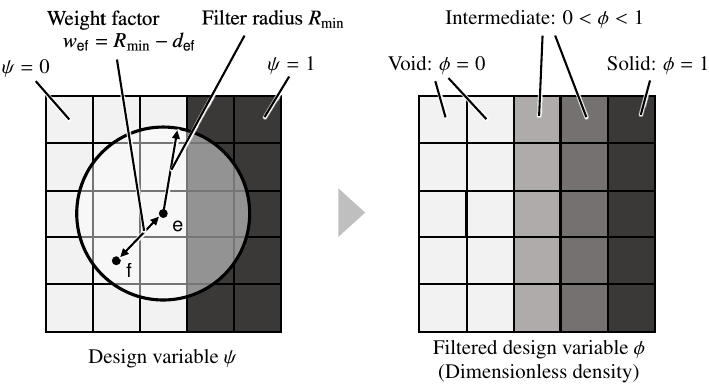}
  \caption{Concept of the filtering process.}
  \label{figfilter}
\end{figure}
\par
Since the design variable $\psi_{\tf}$ can independently take arbitrary values, the optimization process may yield physically meaningless configurations, such as fine-scale patterns.
To mitigate this issue, a filtering process illustrated in \fref{figfilter} is applied to smooth the distribution of the design variables as follows:
\begin{align}\label{eqFilter}
  \phi_{\te} = \frac{\sum_{\tf=1}^{N_{\rm elem}} {w}_{\te\tf} \psi_{\tf}}{\sum_{\tf=1}^{N_{\rm elem}} {w}_{\te\tf}},
\end{align}
where $\phi_{\te}$ is the filtered design variable and ${w}_{\te\tf}$ is the weight factor.
Similarly, these filtered design variables are expressed in vector form as a dimensionless density vector $\bm{\phi} = \{\phi_1, \phi_2, \dots, \phi_{N_{\rm elem}}\}^\top$.
A linear cone kernel is adopted for the weight factor ${w}_{\te\tf}$, which is defined as
\begin{align}\label{eqFilterRadius}
  {w}_{\te\tf} = \max(0, R_{\min} - {d}_{\te\tf}),
\end{align}
where $R_{\min}$ denotes the filter radius and ${d}_{\te\tf}$ is the center-to-center distance between elements $\te$ and $\tf$.
Here, the filtered design variable $\phi_{\te}$ is utilized as the dimensionless material density in the thermal analysis. Specifically, $\phi_{\te} = 1$ represents solid material, $\phi_{\te} = 0$ represents void, and $0 < \phi_{\te} < 1$ represents an intermediate density between solid and void.

\subsection{Governing equations}\label{secGoverningEquation}
This study formulates the conduction-radiation problem within the finite element method (FEM) framework by incorporating radiative heat transfer into the heat source term of the heat conduction equation.
\subsubsection{Heat conduction}\label{secHeatConduction}
\begin{figure}[htbp]
  \begin{center}
    \includegraphics[scale=1.0]{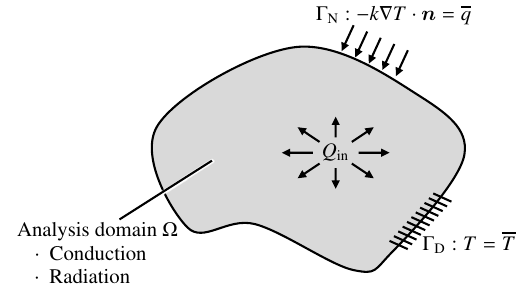}
  \end{center}
  \caption{Analysis domain and boundary conditions of heat conduction analysis.}
  \label{figConductionAnalysis}
\end{figure}
Consider a steady-state conduction-radiation heat transfer problem defined on the analysis domain $\Omega$, as illustrated in \fref{figConductionAnalysis}.
Let $Q_{\rm in}$ and $Q_{\text{rad}}$ denote the volumetric internal heat generation and the volumetric radiative heat emission, respectively.
The governing equations for heat conduction are given by
\begin{align}\label{eqHeatConduction}
  \begin{aligned}
    \underbrace{-\nabla\cdot\left(k(\phi)\nabla T\right)}_{\text{Conduction}} +
    \underbrace{Q_{\text{rad}}}_{\text{Radiation, \eref{eqQrad}}}
     & = Q_{\rm in}
     &              & \text{in} \quad \Omega,         \\
    -k(\phi)\nabla T \cdot\bm{n}
     & = -\bar{q}
     &              & \text{on} \quad \Gamma_{\rm N}, \\
    T
     & = \bar{T}
     &              & \text{on} \quad \Gamma_{\rm D},
  \end{aligned}
\end{align}
where $T$ represents the temperature field within $\Omega$, and $\bm{n}$ is the outward unit normal vector.
The parameters $\bar{q}$ and $\bar{T}$ prescribe the heat flux on the Neumann boundary $\Gamma_{\rm N}$ and the temperature on the Dirichlet boundary $\Gamma_{\rm D}$, respectively, where $\bar{q}$ is taken positive when heat enters the domain.
The temperature field $T$ is primarily defined at the finite element nodes and interpolated within each element using shape functions.
The effective thermal conductivity $k(\phi)$ is defined as a function of the dimensionless material density $\phi$.

\subsubsection{Radiative heat transfer}\label{secRadiativeHeatTransfer}
\begin{figure}[htbp]
  \centering
  \includegraphics[scale=1.0]{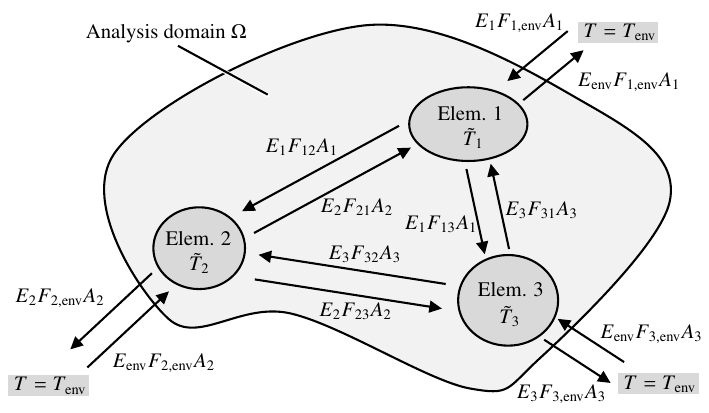}
  \caption{Concept of radiative heat transfer analysis.}
  \label{figRadiationConcept}
\end{figure}
The radiative heat transfer between elements is evaluated using the element-wise representative temperature, as conceptually illustrated in \fref{figRadiationConcept}.
In accordance with the zonal method, each finite element is treated as an isothermal zone \cite{modest2021radiative}.
Let $\tilde{T}_{\te}$ denote the representative temperature of element $\te$ corresponding to that zone.
Assuming that the solid material behaves as a black body and that the void region is a vacuum, the emissive power of element $\te$, denoted by $E_{\te}$, is given by the Stefan-Boltzmann law \cite{modest2021radiative} as follows:
\begin{align}\label{eqemissivePower}
  E_{\te} = \sigma \tilde{T}_{\te}^4,
\end{align}
where $\sigma$ is the Stefan-Boltzmann constant.
In addition to the finite elements, the external environment surrounding the analysis domain is also treated as an isothermal zone denoted by the subscript $\bullet_{\text{env}}$.
This zone is assumed to act as a black body at a prescribed ambient temperature $T_{\text{env}}$, with the emissive power $E_{\text{env}} = \sigma T_{\text{env}}^4$.
The volumetric net radiative heat transfer of element $\te$, denoted by $Q_{{\text{rad}},\te}$, is determined from the balance between the emitted and absorbed radiative energy:
\begin{align}\label{eqQrad}
  \begin{aligned}
    Q_{{\text{rad}}, \te}
     & = \frac{A_{\te} f(\phi_{\te})}{V_{\te}} \left( E_{\te} - \sum_{\re=1}^{N_{\rm elem}} F_{\te\re} E_{\re} - F_{\te,\text{env}} E_{\text{env}} \right)                               \\
     & = \frac{A_{\te} f(\phi_{\te})}{V_{\te}} \sigma \left( \tilde{T}_{\te}^4 - \sum_{\re=1}^{N_{\rm elem}} F_{\te\re} \tilde{T}_{\re}^4 - F_{\te,\text{env}} T_{\text{env}}^4 \right),
  \end{aligned}
\end{align}
in which $A_{\te} f(\phi_{\te})$ represents the effective radiative surface area of element $\te$ with $A_{\te}$ denoting the total surface area of the element $\te$ and $f(\phi_{\te})$ being the density interpolation function satisfying $0 \leq f(\phi_{\te}) \leq 1$.
Also, $V_{\te}$ is the volume of the element $\te$, $F_{\te\re}$ denotes the exchange factor from element $\te$ to element $\re$, and $F_{\te,\text{env}}$ denotes the exchange factor from element $\te$ to the external environment.
These exchange factors, $F_{\te\re}$ and $F_{\te,\text{env}}$, account for not only the geometric relationships between elements but also the attenuation caused by intermediate-density elements.
The calculation procedure for these exchange factors is described in \sref{secExchangeFactor}.

\subsubsection{Nondimensionalization}\label{secNondimensionalization}
To ensure the generality of the formulation, the governing equations are nondimensionalized. Let $L$ denote the characteristic length of the system, $T_{\rm ref}$ denote the reference temperature, and $k_{\rm max}$ denote the thermal conductivity of the solid material used as the reference conductivity. The dimensionless coordinate $\bm{x}^*$, the dimensionless temperature $T^*$, the dimensionless thermal conductivity $k^*(\phi)$, and the dimensionless spatial gradient operator $\nabla^*$ are defined as follows:
\begin{align}\label{eqRepresentativeParameters}
  \bm{x}^* = \frac{\bm{x}}{L}, \quad T^* = \frac{T}{T_{\rm ref}}, \quad k^*(\phi) = \frac{k(\phi)}{k_{\rm max}},\quad \nabla^*=L\nabla.
\end{align}
Substituting these dimensionless variables into \eref{eqHeatConduction} yields the following nondimensionalized governing equation and boundary conditions:
\begin{align}\label{eqnondimGov}
  \begin{aligned}
    -\nabla^*\cdot\left(k^*(\phi)\nabla^* T^*\right) + Q_{\text{rad}}^*
        & = Q_{\rm in}^*
        &                & \text{in} \quad \Omega^*          \\
    -k^*(\phi)\nabla^* T^* \cdot\bm{n}
        & = -\bar{q}^*
        &                & \text{on} \quad \Gamma_{\rm N}^*  \\
    T^* & = \bar{T}^*
        &                & \text{on} \quad \Gamma_{\rm D}^*,
  \end{aligned}
\end{align}
where $Q_{\rm in}^* = Q_{\rm in} L^2 / (k_{\rm max} T_{\rm ref})$ is the dimensionless internal heat generation, $\bar{q}^* = \bar{q} L / (k_{\rm max} T_{\rm ref})$ denotes the dimensionless prescribed heat flux, and $\bar{T}^* = \bar{T} / T_{\rm ref}$ is the dimensionless prescribed temperature.
Additionally, $\Gamma_{\rm N}^*$ and $\Gamma_{\rm D}^*$ denote the dimensionless Neumann and Dirichlet boundaries, respectively, and $\bm{n}$ is the outward unit normal vector.
\par
Furthermore, by introducing the dimensionless surface area of an element $A_{\te}^* = A_{\te} / L^2$ and the dimensionless element volume $V_{\te}^* = V_{\te} / L^3$, the dimensionless radiative heat transfer term $Q_{{\text{rad}}, \te}^*$ is reformulated from \eref{eqQrad} as follows:
\begin{align}\label{eqQradi}
  Q_{{\text{rad}}, \te}^* = N_{\rm R} \frac{A_{\te}^*f(\phi_{\te})}{V_{\te}^*} \left( \tilde{T}_{\te}^{*4} - \sum_{\re=1}^{N_{\rm elem}} F_{\te\re} \tilde{T}_{\re}^{*4} - F_{\te,\text{env}} T_{\text{env}}^{*4}\right),
\end{align}
where $\tilde{T}_{\te}^* = \tilde{T}_{\te} / T_{\rm ref}$ and ${T}_{\rm env}^* = {T}_{\rm env} / T_{\rm ref}$ are the dimensionless representative temperature of element $\te$ and the dimensionless ambient temperature, respectively.
Here, $N_{\rm R}$ is the conduction-radiation parameter \cite{talukdar2002analysis,cintolesi2017numerical} defined as
\begin{align}\label{eqNR}
  N_{\rm R} = \frac{\sigma T_{\rm ref}^3 L}{k_{\rm max}}.
\end{align}
This parameter characterizes the relative magnitude of radiative heat transfer with respect to conductive heat transfer within the system.
Consequently, the system is entirely governed by the conduction-radiation parameter $N_{\rm R}$ and the dimensionless parameters prescribing the thermal conditions ($Q^*_{\rm in}$, $\bar{q}^*$, and $\bar{T}^*$).

\subsection{Exchange factors}\label{secExchangeFactor}
This subsection details the evaluation of the exchange factor $F_{\te\re}$, which represents the fraction of radiative energy leaving element $\te$ that is absorbed by element $\re$.
To evaluate the attenuation effects caused by intermediate-density materials, this study adopts a ray-tracing approach \cite{lockwood1981new,modest2021radiative}.
\sref{secViewFactor} describes the calculation of view factors without attenuation by discretizing the surface area and solid angles.
Subsequently, \sref{secShadowingEffects} formulates the exchange factors that account for actual attenuation effects by evaluating ray attenuation through the elements.
\subsubsection{View factors}\label{secViewFactor}
\begin{figure}[htbp]
  \begin{center}
    \includegraphics[scale=1.0]{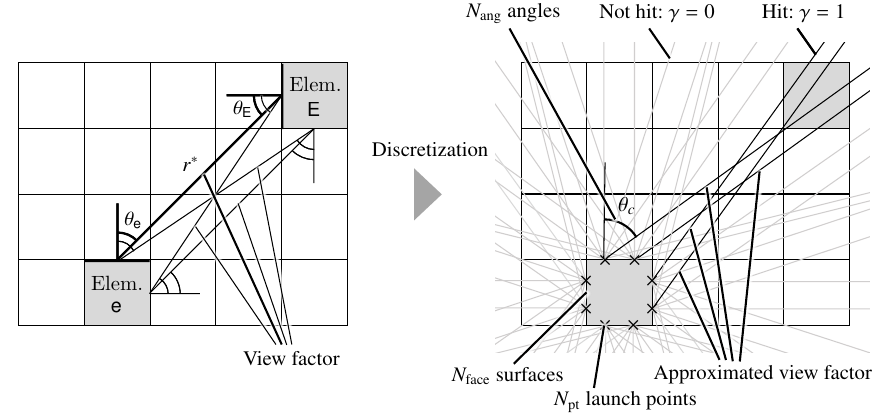}
  \end{center}
  \caption{Concept of the view factor evaluation using a ray tracing.}
  \label{figRayTracing}
\end{figure}
The view factor $F^{\text{geo}}_{\te\re}$, which represents the purely geometric relationship between two surfaces, is formulated without considering attenuation effects between two elements.
Let $\Gamma_{\te}^*$ and $\Gamma_{\re}^*$ denote the dimensionless geometric surfaces of elements $\te$ and $\re$, respectively. The view factor from element $\te$ to element $\re$ is analytically defined as follows:
\begin{align}\label{eqFijgeo}
  \begin{aligned}
    F^{\text{geo}}_{\te\re}
     & = \frac{1}{A_{\te}^*}\int_{\Gamma_{\te}^*}\int_{\Gamma_{\re}^*}\frac{\langle\cos\theta_{\te}\rangle \langle\cos\theta_{\re}\rangle}{\pi r^{*2}}\ \text{d}A_{\re}^* \ \text{d}A_{\te}^* \\
     & = \frac{1}{A_{\te}^*}\int_{\Gamma_{\te}^*}\int_{\omega_{\re}}\frac{\langle\cos\theta_{\te}\rangle}{\pi }\ \text{d}\omega_{\re}\ \text{d}A_{\te}^*,
  \end{aligned}
\end{align}
where $\theta_{\te}$ and $\theta_{\re}$ denote the angles between the surface normals and the line connecting the infinitesimal areas $\text{d}A_{\re}^* $ and $ \text{d}A_{\te}^* $, and $r^*$ denotes the dimensionless distance between the infinitesimal areas.
Here, $\text{d}\omega_{\re} = \left(\langle\cos\theta_{\re}\rangle / r^{*2}\right)\text{d}A_{\re}^*$ denotes the infinitesimal solid angle of $\text{d}A_{\re}^*$ viewed from $\text{d}A_{\te}^*$.
Notably, the Macaulay bracket $\langle\,\bullet\,\rangle = \max(\,\bullet\,,0)$ excludes surface pairs facing away from each other.
\par
To numerically evaluate the view factor, the continuous integration is approximated through spatial and angular discretizations, as illustrated in \fref{figRayTracing}.
For an element consisting of $N_{\rm face}$ faces with a dimensionless geometric area of $A_{\te}^{a*}$ for each face, the surface is discretized into $N_{\rm pt}$ sections, and a launch point is placed at the center of each section.
The angular space above each launch point is further discretized using a ray-tracing approach.
In three-dimensional problems, the azimuthal angle and the cosine of the zenith angle are uniformly discretized over the hemisphere to achieve a uniform solid-angle distribution, whereas in two-dimensional problems the in-plane angle is uniformly discretized over the half-plane.
In both cases, a total of $N_{\rm ang}$ rays are emitted from each launch point.
Based on this discretization, the view factor $F^{\mathrm{geo}}_{\te\re}$ can be evaluated by
\begin{align}\label{eqGeometricViewFactor}
  F^{\text{geo}}_{\te\re} \approx \frac{1}{A_{\te}^*} \sum_{a=1}^{N_{\rm face}} \sum_{b=1}^{N_{\rm pt}} \sum_{c=1}^{N_{\rm ang}} \frac{\mathcal{C} A_{\te}^{a*} \gamma_{a,b,c}^{(\re)} \cos\theta_c}{N_{\rm pt}},
\end{align}
where $\mathcal{C}$ is the normalization coefficient for the angular discretization.
It is defined as $\mathcal{C} = \pi / (2N_{\rm ang})$ for two-dimensional problems and $\mathcal{C} = 2 / N_{\rm ang}$ for three-dimensional problems.
The variable $\gamma_{a,b,c}^{(\re)}$ is a Boolean indicator that takes a value of $1$ when the $c$-th ray emitted from the $b$-th launch point on surface $a$ intersects element $\re$, and $0$ otherwise.

\subsubsection{Attenuation effects}\label{secShadowingEffects}
\begin{figure}[htbp]
  \begin{center}
    \includegraphics[scale=0.9]{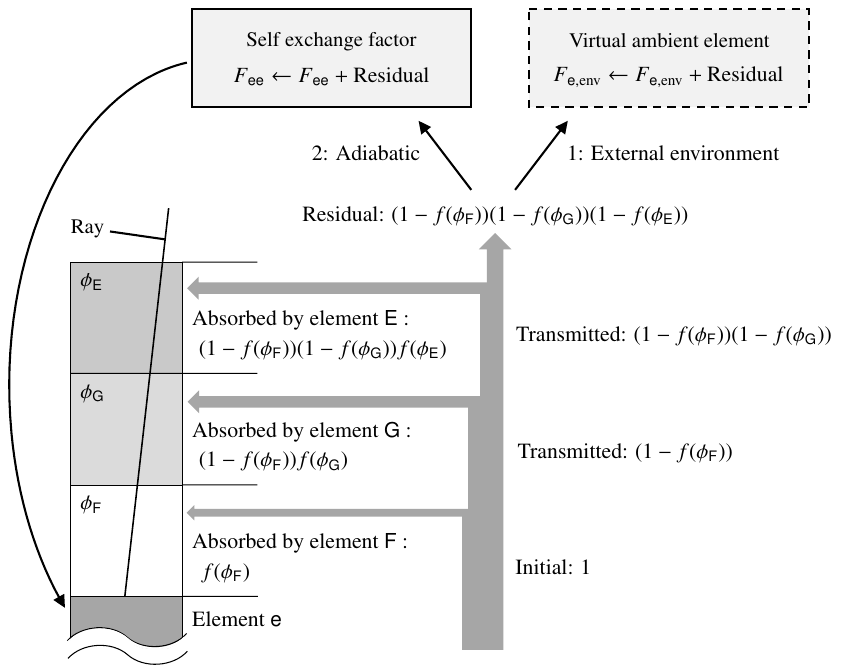}
  \end{center}
  \caption{Calculation of the attenuation effect.}
  \label{figShadowing}
\end{figure}
During the TO process, regions with intermediate densities, $0 < \phi < 1$, inevitably emerge.
This study interprets these intermediate-density regions as porous microstructures composed of a mixture of opaque solid material and void.
When a ray passes through these regions, a portion of its radiative energy is attenuated due to absorption by the internal solid surfaces of the microstructures.
Based on Kirchhoff's law of thermal radiation, the absorptivity of the intermediate material must be equal to its effective emissivity to maintain thermodynamic consistency.
Hence, the absorptivity of the microstructures is evaluated using the same density interpolation function $f(\phi)$ introduced in \eref{eqQrad}.
\par
\fref{figShadowing} illustrates the attenuation process of a ray propagating through multiple elements.
In the present framework, the absorptivity and transmissivity of a ray through an intermediate element are defined as $f(\phi)$ and $1-f(\phi)$, respectively.
Specifically, consider a ray emitted from the reference element $\te$ toward a target element $\re$.
If the ray passes through intermediate elements $\rf$ and $\rg$ before arriving at element $\re$, the cumulative transmissivity associated with the ray is evaluated as $\left(1-f(\phi_{\rf})\right)\left(1-f(\phi_{\rg})\right)$, thus the fraction of the emitted energy absorbed by element $\re$ is $\left(1-f(\phi_{\rf})\right)\left(1-f(\phi_{\rg})\right)f(\phi_{\re})$.
Accordingly, the exchange factor $F_{\te\re}$ is formulated by incorporating the attenuation effects into the ray-tracing evaluation of the geometric view factor $F^{\text{geo}}_{\te\re}$.
Specifically, the geometric contribution of each ray is multiplied by its cumulative transmissivity along the ray propagation path and the absorption rate within element $\re$ itself, as follows:
\begin{align}\label{eqExchangeFactor}
  F_{\te\re} \approx
  \frac{1}{A_{\te}^*} \sum_{a=1}^{N_{\rm face}} \sum_{b=1}^{N_{\rm pt}} \sum_{c=1}^{N_{\rm ang}}
  \frac{\mathcal{C} A_{\te}^{a*} \gamma_{a,b,c}^{(\re)} \cos\theta_c}{N_{\rm pt}}
  \left\{ \prod_{d=1}^{N_{\rm elem}} \left(1-{H}_{\te\re}^{d}f(\phi_{d})\right) \right\} f(\phi_{\re}),
\end{align}
in which the product term $\prod_{d=1}^{N_{\rm elem}} \left(1-{H}_{\te\re}^{d}f(\phi_{d})\right)$ represents the remaining fraction of radiative energy immediately before reaching element $\re$. Also, ${H}_{\te\re}^{d}$ is a binary indicator function defined as
\begin{align}\label{eqray_indicator}
  {H}_{\te\re}^{d}=
  \begin{cases}
    1, & \text{if element $d$ lies on the ray propagation path between elements $\te$ and $\re$}, \\
    0, & \text{otherwise}.
  \end{cases}
\end{align}
Additionally, the final term $f(\phi_{\re})$ represents the fraction of energy absorbed as heat by element $\re$.
Note that, in ray-tracing approaches, incorporating attenuation based on the ray path length generally requires either a prohibitively large number of rays or complex geometric corrections \cite{coelho1997conservative} to rigorously satisfy Kirchhoff's law.
Furthermore, an attenuation model based on the ray path length erroneously permits the transmission of rays through thin solid structures.
To avoid these issues and ensure thermodynamic consistency without excessive computational cost, the present study introduces an artificial attenuation model that neglects the ray path length and represents the attenuation solely through the density interpolation function $f(\phi)$.
Although this treatment introduces a mesh dependency regarding the attenuation in intermediate density regions, it does not deteriorate the validity of the calculation results.
This is because an implicit surface consisting of intermediate densities exhibits the same radiation behavior as an explicit boundary (see \sref{subsecHeatSink2D}).
\par
Meanwhile, the radiative energy that escapes the analysis domain without being absorbed by the structure must be properly accounted for to ensure satisfaction of the macroscopic thermal boundary conditions.
For a ray leaving the analysis domain, the residual energy fraction $R$ associated with the ray can be calculated by
\begin{align}\label{eqRayResidual}
  R = \frac{1}{A_{\te}^*}\frac{\mathcal{C} A_{\te}^{a*} \cos\theta_c}{N_{\rm pt}}
  \left\{ \prod_{{d}=1}^{{N_{\rm elem}}} \left(1-{{H}_{\te,{\text{env}}}^{d}}f(\phi_{{d}})\right) \right\},
\end{align}
in which ${{H}_{\te,{\text{env}}}^{d}}$ is a binary indicator function defined in the same manner as \eref{eqray_indicator}.
Specifically, it takes the value of 1 if element $d$ lies on the propagation path of a ray escaping from the reference element $\te$ to the external environment, and 0 otherwise.
Depending on the thermal boundary condition associated with the ray escape direction, the residual energy fraction $R$ is incorporated into the exchange factors as follows:
\begin{itemize}
  \item
        If a ray exits the domain toward an external environment with a prescribed ambient temperature, the residual energy fraction is assigned to the exchange factor of the virtual ambient element, $F_{\te,\text{env}}$:
        \begin{align}\label{eqUpdateEnv}
          F_{\te, \text{env}} \leftarrow F_{\te, \text{env}} + R.
        \end{align}
  \item
        If a ray exits the domain toward an adiabatic direction, the residual energy is assigned to the self-exchange factor $F_{\te\te}$ of the emitting element to satisfy the zero-flux condition:
        \begin{align}\label{eqUpdateSelf}
          F_{\te\te} \leftarrow F_{\te\te} + R.
        \end{align}
\end{itemize}
This treatment ensures the conservation of the total emitted radiative energy within the global system.

\subsection{Discretization}\label{secDiscretization}
Applying finite element discretization to the dimensionless governing equations in \eref{eqnondimGov} yields a system of nonlinear algebraic equations, in which the dimensionless nodal temperature vector $\mathbf{T}^* \in \mathbb{R}^{N_{\rm node}}$ is the unknown variable.
The residual vector $\mathbf{R} \in \mathbb{R}^{N_{\rm node}}$ of the system is defined as follows:
\begin{align}\label{eqResidual}
  \mathbf{R}
   & = \mathbf{K}_{\rm cond}^*\mathbf{T}^*
  +\mathbf{F}_{\text{rad}}^*
  -\mathbf{F}_{\rm in}^* -\mathbf{F}_{{\rm N}}^* = \mathbf{0},
\end{align}
where $\mathbf{K}_{\rm cond}^* \in \mathbb{R}^{N_{\rm node} \times N_{\rm node}}$ is the global thermal conductivity matrix,  $\mathbf{F}_{\text{rad}}^* \in \mathbb{R}^{N_{\rm node}}$ is the radiative heat emission vector, $\mathbf{F}_{\rm in}^* \in \mathbb{R}^{N_{\rm node}}$ is the internal heat generation vector, and $\mathbf{F}_{\rm N}^* \in \mathbb{R}^{N_{\rm node}}$ is the heat flux vector arising from the Neumann boundary conditions, respectively.
These matrices and vectors are assembled from the element contributions as
\begin{align}\label{eqResidual_comp}
  \mathbf{K}_{\rm cond}^*
   & = \bigcup_{\te=1}^{N_{\rm elem}} \int_{\Omega^{\te*}} \mathbf{B}_{\te}^\top k^*(\phi_{\te}) \mathbf{B}_{\te} \,\text{d}\Omega^*,                             \\
  \mathbf{F}_{\text{rad}}^*
   & = \bigcup_{\te=1}^{N_{\rm elem}} \int_{\Omega^{\te*}} \mathbf{N}_{\te}^\top \underbrace{Q_{{\text{rad}, \te}}^*}_{\text{\eref{eqQradi}}} \,\text{d}\Omega^*, \\
  \mathbf{F}_{\rm in}^*
   & = \bigcup_{\te=1}^{N_{\rm elem}} \int_{\Omega^{\te*}} \mathbf{N}_{\te}^\top Q_{\rm in}^* \,\text{d}\Omega^*,                                                 \\
  \mathbf{F}_{{\rm N}}^*
   & = \bigcup_{\te=1}^{N_{\rm elem}} \int_{\Gamma_{\rm N}^{\te*}} \mathbf{N}_{\te}^\top \bar{q}^* \,\text{d}\Gamma^*,
\end{align}
where $\bigcup_{\te=1}^{N_{\rm elem}}$ denotes the global assembly operator, and $\Omega^{\te*}$ and $\Gamma_{\rm N}^{\te*}$ represent the dimensionless volume and Neumann boundary of element $\te$, respectively.
The matrices $\mathbf{N}_{\te}$ and $\mathbf{B}_{\te}$ are the element-level shape function vector and its gradient matrix, respectively.
Note that the Dirichlet boundary condition $T^* = \bar{T}^*$ on $\Gamma_{\rm D}^*$ is enforced on the global system of equations prior to the solution process.
As explained before, the volumetric radiative heat emission $Q_{\text{rad},\te}^*$ depends on the representative temperature of element $\te$. In the present study, the dimensionless representative temperature of element $\te$ is defined as its volume-averaged temperature, given by
\begin{align}\label{eqRepresentativeTemperature}
  \tilde{T}_{\te}^* = \frac{1}{V_{\te}^{*}}\int_{\Omega^{\te*}}\mathbf{N}_{\te} \mathbf{T}_{\te}^{*} \, \text{d}\Omega^{*},
\end{align}
where $\mathbf{T}_{\te}^{*}$ denotes the element-level dimensionless temperature vector.
\par
To solve the nonlinear equation presented in \eref{eqResidual} using the Newton-Raphson method, the following iterative procedure is performed:
\begin{align}\label{eqNewtonRaphson}
  \mathbf{K}_{\rm tan}^* \Delta\mathbf{T}^* & = -\mathbf{R},
\end{align}
where $\mathbf{K}_{\rm tan}^* \in \mathbb{R}^{N_{\rm node} \times N_{\rm node}}$ is the global tangent stiffness matrix.
This matrix is derived by differentiating the residual vector $\mathbf{R}$ with respect to the nodal temperature vector $\mathbf{T}^*$ as follows:
\begin{align}\label{eqKtan}
  \begin{aligned}
    \mathbf{K}_{\rm tan}^*
     & = \frac{\partial \mathbf{R}}{\partial \mathbf{T}^*}
    \\
     & = \underbrace{\mathbf{K}_{\rm cond}^*}_{\text{\eref{eqResidual_comp}}}
    +\mathbf{K}_{\text{rad}}^*.
  \end{aligned}
\end{align}
In the above expression, $\mathbf{K}_{\text{rad}}^* \in \mathbb{R}^{N_{\rm node} \times N_{\rm node}}$ is the tangent stiffness matrix associated with the radiative heat emission vector.
By differentiating $\mathbf{F}_{\text{rad}}^*$ and substituting the formulation of $Q_{{\text{rad}}, \te}^*$ defined in \eref{eqQradi}, this matrix is derived as follows:
\begin{align}\label{eqKrad}
  \begin{aligned}
    \mathbf{K}_{{\text{rad}}}^*
     & = \frac{\partial \mathbf{F}_{\text{rad}}^*}{\partial \mathbf{T}^*}                                                                                                                                                                                      \\
     & = \frac{\partial}{\partial \mathbf{T}^*} \bigg(\bigcup_{\te=1}^{N_{\rm elem}} \int_{\Omega^{\te*}} \mathbf{N}_{\te}^\top  Q_{{\text{rad}}, \te}^* \, \text{d}\Omega^* \bigg)                                                                            \\
     & = \bigcup_{\te=1}^{N_{\rm elem}} \int_{\Omega^{\te*}} \mathbf{N}_{\te}^\top N_{\rm R} \frac{A_{\te}^*f(\phi_{\te})}{V_{\te}^*} \left( 4\tilde{T}_{\te}^{*3}\right) \mathbf{M}_{\te} \, \text{d}\Omega^*                                                 \\
     & \quad - \bigcup_{\te=1}^{N_{\rm elem}}\bigcup_{\re=1}^{N_{\rm elem}} \int_{\Omega^{\te*}} \mathbf{N}_{\te}^\top N_{\rm R} \frac{A_{\te}^*f(\phi_{\te})}{V_{\te}^*} \left( 4 F_{\te\re} \tilde{T}_{\re}^{*3}\right)\mathbf{M}_{\re} \, \text{d}\Omega^*.
  \end{aligned}
\end{align}
Here, $\mathbf{M}_{\te}$ and $\mathbf{M}_{\re}$ denote the element averaging vectors, defined as
\begin{align}\label{eqeleave}
  \begin{aligned}
    \mathbf{M}_{\te} &
    = \frac{1}{V_{\te}^*}\int_{\Omega^{\te*}} \mathbf{N}_{\te} \, \text{d}\Omega^*
    = \frac{1}{V_{\te}^*}\left[ \int_{\Omega^{\te*}} N_{\te,1} \, \text{d}\Omega^*, \ \int_{\Omega^{\te*}} N_{\te,2} \, \text{d}\Omega^*, \ \dots, \ \int_{\Omega^{\te*}} N_{\te, N_{\rm npe}} \, \text{d}\Omega^* \right],
    \\
    \mathbf{M}_{\re} &
    = \frac{1}{V_{\re}^*}\int_{\Omega^{\re*}} \mathbf{N}_{\re} \, \text{d}\Omega^*
    = \frac{1}{V_{\re}^*}\left[ \int_{\Omega^{\re*}} N_{\re,1} \, \text{d}\Omega^*, \ \int_{\Omega^{\re*}} N_{\re,2} \, \text{d}\Omega^*, \ \dots, \ \int_{\Omega^{\re*}} N_{\re, N_{\rm npe}} \, \text{d}\Omega^* \right],
  \end{aligned}
\end{align}
where $N_{\te,k}$ and $N_{\re,k}$ are the shape functions associated with the $k$-th local node of elements $\te$ and $\re$, and $N_{\rm npe}$ is the number of nodes per element.
As can be seen from \eref{eqKrad} and \eref{eqeleave}, the evaluation of radiative interactions involves numerical integrations between remote elements. Consequently, the resulting global tangent matrix is no longer strictly sparse, but becomes considerably denser than that of conventional finite element formulations based solely on local interactions.
\par
To verify the accuracy of the proposed method, several benchmark simulations are presented in \saref{appVerificationSimulation}.
The numerical results show agreement with reference solutions, demonstrating the reliability of the forward analysis.

\subsection{Topology optimization}\label{secTopologyOptimization}
This subsection presents the formulation of the TO. First, the material interpolation schemes for heat conduction and radiation are introduced in \sref{secMaterialInterpolation}. Next, the optimization problem is formulated in \sref{secOptimizationProblem}.
Subsequently, the sensitivity of the objective function is derived using the adjoint variable method in \sref{secSensitivityAnalysis}.
Finally, the overall optimization algorithm based on these formulations is described in \sref{secOptimizationAlgorithm}.
\subsubsection{Material interpolation scheme}\label{secMaterialInterpolation}
In this study, a density-based approach is adopted to parameterize the material properties based on the dimensionless material density $\phi$ defined in \sref{secDesignVariables}. To suppress the formation of intermediate densities, the Solid Isotropic Material with Penalization (SIMP) method is applied to both the dimensionless thermal conductivity $k^*(\phi)$ introduced in \eref{eqRepresentativeParameters} and thermal radiation interpolation $f(\phi)$ introduced in \eref{eqQrad}:
\begin{align}\label{eqInterpolationk}
  k^*(\phi) & = k_{\min}^* + (1 - k_{\min}^*) \phi^p,
  \\ \label{eqInterpolationphi}
  f(\phi)   & = \phi^q,
\end{align}
where $p$ and $q$ are the penalization parameters for thermal conduction and radiation, respectively, and $k_{\min}^*$ is a small lower bound value introduced for numerical stability.
\par
The choice of these penalization parameters significantly affects the optimized structure by determining the effectiveness of intermediate densities for thermal conduction and radiation relative to their volume cost.
A detailed discussion regarding the effects of these interpolation schemes on the optimization results is provided in \saref{appInterpolation}.

\subsubsection{Optimization problem}\label{secOptimizationProblem}
The objective of the optimization is to improve radiative heat dissipation while maintaining a lightweight structure.
In this formulation, the entire analysis domain $\Omega^*$ is defined as the design domain.
Accordingly, the temperature integrated over the target evaluation domain $\Omega_{\rm obj}^*$ is minimized subject to a volume constraint imposed on the entire design domain.
Using the discrete system derived in \sref{secDiscretization}, the optimization problem is defined as follows:
\begin{align}\label{eqoptproblem}
  \begin{aligned}
    \text{Find} \quad       &
    \bm{\psi} = \{\psi_1, \psi_2, \dots, \psi_{N_{\rm elem}}\}^\top                                                    , \\
    \text{Minimize} \quad   &
    J= \sum_{\te \in \mathcal{N}_{\rm obj}} \int_{\Omega^{\te*}} \mathbf{N}_{\te} \mathbf{T}_{\te}^* \text{d}\Omega^*
    ,                                                                                                                    \\
    \text{Subject to} \quad &
    \left\{~~
    \begin{aligned}
       & \text{Filtering:} \quad \text{\eref{eqFilter}}                           \\
       & \text{Governing equation:}  \quad \text{\eref{eqResidual}}               \\
       & V^* = \sum_{{\te}=1}^{N_{\rm elem}} \phi_{\te} V_{\te}^* \leq V_{\max}^* \\
       & 0 \leq \psi_{\tf} \leq 1 \quad ({\tf} = 1, 2, \dots, N_{\rm elem})
    \end{aligned}
    \right.,
  \end{aligned}
\end{align}
where $J$ is the objective function, and $\bm{\psi}$ is the vector of design variables.
The index set $\mathcal{N}_{\rm obj}$ contains the indices of elements belonging to the target evaluation domain $\Omega_{\rm obj}^*$.
$V^*$ represents the total dimensionless volume of the solid material, and $V_{\max}^*$ is the prescribed upper limit of the dimensionless volume.

\subsubsection{Adjoint sensitivity analysis}\label{secSensitivityAnalysis}
In this study, the sensitivity of the objective function with respect to the filtered design variable $\phi_{\de}$ is derived using the adjoint variable method.
First, the Lagrangian $\mathcal{L}$, formulated by adding the residual vector of the governing equations to the objective function, is defined as follows:
\begin{align}\label{eqlag}
  \begin{aligned}
    \mathcal{L}
     & = J + \bm{\lambda}^\top \mathbf{R}\left(\bm{\phi}, \mathbf{T}^*\right),
  \end{aligned}
\end{align}
where $\bm{\lambda} \in \mathbb{R}^{N_{\rm node}}$ is an arbitrary adjoint variable vector.
Since the governing equation $\mathbf{R} = \mathbf{0}$ is strictly satisfied during the forward analysis, the Lagrangian $\mathcal{L}$ is equivalent to the original objective function $J$.
Taking the total derivative of $\mathcal{L}$ with respect to the dimensionless density $\phi_{\de}$ yields
\begin{align}\label{eqTotalDerivative}
  \frac{\text{d} \mathcal{L}}{\text{d} \phi_{\de}}
  = \frac{\partial \mathcal{L}}{\partial \phi_{\de}} + \frac{\partial \mathcal{L}}{\partial \mathbf{T}^*} \frac{\text{d} \mathbf{T}^*}{\text{d} \phi_{\de}}.
\end{align}
To eliminate the computationally expensive implicit derivative term $\text{d}\mathbf{T}^*/\text{d}\phi_{\de}$, the adjoint vector $\bm{\lambda}$ is chosen such that the partial derivative of the Lagrangian with respect to the state variable $\mathbf{T}^*$ vanishes
\begin{align}\label{eqdLdT}
  \frac{\partial \mathcal{L}}{\partial \mathbf{T}^*} =
  \frac{\partial J}{\partial \mathbf{T}^*} + \bm{\lambda}^\top
  \underbrace{\mathbf{K}_{\rm tan}^*}_{\text{\eref{eqKtan}}} = \mathbf{0}.
\end{align}
Transposing this stationarity condition gives the following adjoint equation for $\bm{\lambda}$:
\begin{align}\label{eqAdjointEq}
  \mathbf{K}_{\rm tan}^{*\top} \bm{\lambda} = - \left( \frac{\partial J}{\partial \mathbf{T}^*} \right)^\top.
\end{align}
\par
If \eref{eqAdjointEq} is satisfied, the second term on the right-hand side of \eref{eqTotalDerivative} becomes zero.
Consequently, the total derivative of the objective function simply equals the explicit partial derivative of the Lagrangian:
\begin{align}\label{eqSensitivityPhi}
  \frac{\text{d} J}{\text{d} \phi_{\de}}
  = \frac{\partial \mathcal{L}}{\partial \phi_{\de}}
  =\underbrace{\frac{\partial J}{\partial \phi_{\de}}}_{=0 \,(\text{See \eref{eqoptproblem}})}
  +\bm{\lambda}^\top \frac{\partial \mathbf{R}}{\partial \phi_{\de}}.
\end{align}
By focusing on the partial derivatives of the thermal conductivity matrix and the radiation-associated terms, \eref{eqSensitivityPhi} is expanded using the chain rule as follows:
\begin{align}\label{eqdJdphi}
  \begin{aligned}
    \frac{\text{d} J}{\text{d} \phi_{\de}}
     & = \underbrace{\bm{\lambda}^\top \frac{\partial \mathbf{K}^*_{\rm cond}}{\partial \phi_{\de}}\mathbf{T}^*}_{\text{Conduction term}}
    + \underbrace{\bm{\lambda}^\top\frac{\partial \mathbf{F}^*_{\text{rad}}}{\partial \phi_{\de}}}_{\text{Radiation term}},
  \end{aligned}
\end{align}
where
\begin{align}\label{eqdKdphi}
  \frac{\partial \mathbf{K}_{\rm cond}^*}{\partial \phi_{\de}}
   & = \bigcup_{\te=1}^{N_{\rm elem}} \int_{\Omega^{\te*}} \mathbf{B}_{\te}^\top \frac{\partial k^*(\phi_{\te})}{\partial \phi_{\de}} \mathbf{B}_{\te} \text{d}\Omega^*,
\end{align}
\begin{align}\label{eqdFraddphi}
  \begin{aligned}
    \frac{\partial \mathbf{F}_{\text{rad}}^*}{\partial \phi_{\de}}
     & = \bigcup_{\te=1}^{N_{\rm elem}} \int_{\Omega^{\te*}} \mathbf{N}_{\te}^\top
    N_{\rm R} \frac{A_{\te}^*}{V_{\te}^*}\frac{\partial f(\phi_{\te})}{\partial \phi_{\de}}  \tilde{T}_{\te}^{*4}
    \, \text{d}\Omega^*
    \\
     & \quad - \bigcup_{\te=1}^{N_{\rm elem}} \bigcup_{\re=1}^{N_{\rm elem}}
    \int_{\Omega^{\te*}} \mathbf{N}_{\te}^\top
    N_{\rm R} \frac{A_{\te}^*}{V_{\te}^*}
    \bigg(
    \frac{\partial f(\phi_{\te})}{\partial \phi_{\de}} F_{\te\re}
    +f(\phi_{\te})\frac{\partial F_{\te\re}}{\partial \phi_{\de}}
    \bigg)\tilde{T}_{\re}^{*4}
    \, \text{d}\Omega^*
    \\
     & \quad - \bigcup_{\te=1}^{N_{\rm elem}}
    \int_{\Omega^{\te*}} \mathbf{N}_{\te}^\top
    N_{\rm R} \frac{A_{\te}^*}{V_{\te}^*}
    \bigg(
    \frac{\partial f(\phi_{\te})}{\partial \phi_{\de}} F_{\te,\text{env}}
    +f(\phi_{\te})\frac{\partial F_{\te,\text{env}}}{\partial \phi_{\de}}
    \bigg)T_{\text{env}}^{*4}
    \, \text{d}\Omega^*,
  \end{aligned}
\end{align}
in which the derivative ${\partial F_{\te\re}}/{\partial \phi_{\de}}$ for $\te \neq \re$ is obtained by differentiating the exchange factor in \eref{eqExchangeFactor} with respect to the density $\phi_{\de}$ as
\begin{align}\label{eqExchangeFactor_derivative}
  \begin{aligned}
    \frac{\partial F_{\te\re}}{\partial \phi_{\de}}
     & =\frac{1}{A_{\te}^*} \sum_{a=1}^{N_{\rm face}} \sum_{b=1}^{N_{\rm pt}} \sum_{c=1}^{N_{\rm ang}}
    \frac{\mathcal{C} A_{\te}^{a*} \gamma_{a,b,c}^{(\re)} \cos\theta_c}{N_{\rm pt}}
    \\
     & \quad \times \bigg[
    -H_{\te\re}^{\de}\frac{\mathrm{d} f(\phi_{\de})}{\mathrm{d}\phi_{\de}}
    \bigg\{ \prod_{\substack{d=1                                                                       \\ d\neq\de}}^{N_{\rm elem}} \left(1-H_{\te\re}^{d}f(\phi_{d})\right) \bigg\} f(\phi_{\re})
    \\&\qquad
    + \bigg\{ \prod_{d=1}^{N_{\rm elem}} \left(1-H_{\te\re}^{d}f(\phi_{d})\right) \bigg\}
    \frac{\partial f(\phi_{\re})}{\partial \phi_{\de}}
    \bigg]\quad (\te \neq\re).
  \end{aligned}
\end{align}
For a ray that escapes the analysis domain without being absorbed, the corresponding residual energy fraction $R$ in \eref{eqRayResidual} also depends on the density field through its path transmissivity.
Its derivative is calculated as
\begin{align}\label{eqResidual_derivative}
  \frac{\partial R}{\partial \phi_{\de}}
  = \frac{1}{A_{\te}^*}
  \frac{\mathcal{C} A_{\te}^{a*}\cos\theta_c}{N_{\rm pt}}
  \bigg(
  -H_{\te,\text{env}}^{\de}\frac{\mathrm{d} f(\phi_{\de})}{\mathrm{d}\phi_{\de}}
  \prod_{\substack{d=1 \\ d\neq\de}}^{N_{\rm elem}}\!\left(1-H_{\te,\text{env}}^{d}f(\phi_{d})\right)
  \bigg).
\end{align}
In the same manner as the forward analysis in \erefs{eqUpdateEnv} and \eqref{eqUpdateSelf}, this derivative is accumulated into the exchange factor derivative corresponding to the boundary condition associated with the escape direction of the ray.
If the ray escapes toward the external environment, it contributes to the environment term:
\begin{align}\label{eqUpdateEnv_derivative}
  \frac{\partial F_{\te,\text{env}}}{\partial \phi_{\de}}
  \leftarrow
  \frac{\partial F_{\te,\text{env}}}{\partial \phi_{\de}} + \frac{\partial R}{\partial \phi_{\de}}.
\end{align}
If the ray escapes toward an adiabatic direction, it contributes to the self-exchange term:
\begin{align}\label{eqUpdateSelf_derivative}
  \frac{\partial F_{\te\te}}{\partial \phi_{\de}}
  \leftarrow
  \frac{\partial F_{\te\te}}{\partial \phi_{\de}} + \frac{\partial R}{\partial \phi_{\de}}.
\end{align}
Note that this evaluation requires a backward traversal of each ray propagation path to collect the downstream elements whose exchange factors are affected by the attenuation change at a target element.
\par
Finally, the chain rule based on density filtering is applied to derive the total sensitivity of the objective function with respect to the original design variable $\psi_{\df}$.
From the definition of the density filter described in \eref{eqFilter}, the partial derivative of the filtered density $\phi_{\de}$ with respect to the design variable $\psi_{\df}$ is obtained as follows:
\begin{align}\label{eqdphidpsi}
  \frac{\partial \phi_{\de}}{\partial \psi_{\df}} =
  \frac{{w}_{\de\df}}{\sum_{\dg=1}^{N_{\rm elem}} {w}_{\de\dg}}.
\end{align}
Consequently, the final sensitivity of the objective function with respect to the design variable $\psi_{\df}$ is evaluated by
\begin{align}\label{eqSensitivity}
  \frac{\text{d} J}{\text{d} \psi_{\df}} =
  \sum_{\de=1}^{N_{\rm elem}}
  \frac{\text{d} J}{\text{d} \phi_{\de}} \frac{\partial \phi_{\de}}{\partial \psi_{\df}}.
\end{align}
The accuracy of the adjoint sensitivity is numerically verified in \saref{appVerificationSensitivity}.

\subsubsection{Optimization algorithm}\label{secOptimizationAlgorithm}
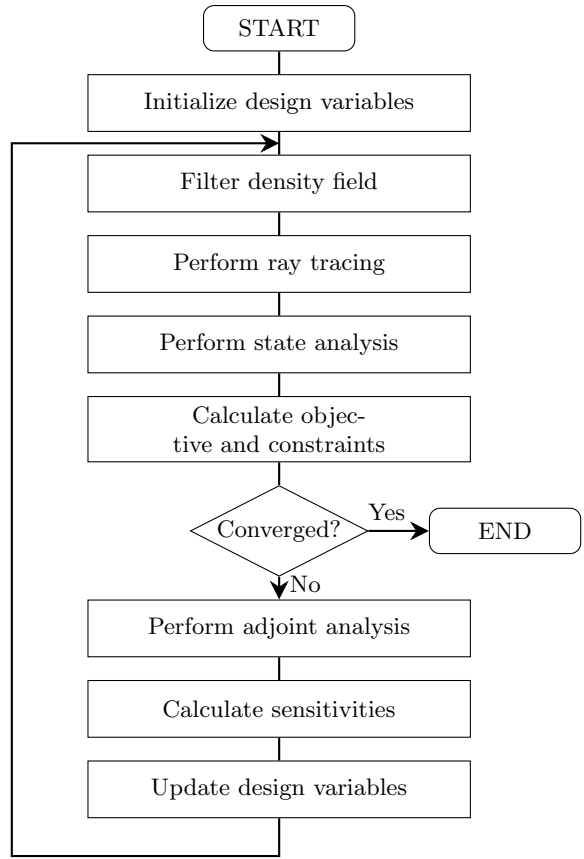
\begin{figure}[htbp]
  \centering
  \begin{tikzpicture}[node distance=3mm and 8mm, font=\footnotesize,color=black]
    \tikzset{
    StartEnd/.style = {rectangle, rounded corners, draw, minimum height=6mm, minimum width=20mm, text centered, fill=white},
    Process/.style  = {rectangle, draw, minimum height=7.5mm, text width=48mm, text centered, fill=white},
    Decision/.style = {diamond, draw, aspect=2.5, minimum height=12mm, inner sep=0pt, text centered, fill=white},
    Arrow/.style    = {thick, -{Stealth[length=2.5mm, width=2.5mm]}},
    Line/.style     = {thick} 
    }
    \node[StartEnd] (start) {START};
    \node[Process, below=of start] (init) {Initialize design variables};
    \node[Process, below=of init] (filter) {Filter density field};
    \node[Process, below=of filter] (ray tracing) {Perform ray tracing};
    \node[Process, below=of ray tracing] (state) {Perform state analysis};
    \node[Process, below=of state] (obj) {Calculate objective and constraints};
    \node[Decision, below=of obj] (conv) {Converged?};
    \node[StartEnd, right=of conv] (end) {END};
    \node[Process, below=of conv] (adjoint) {Perform adjoint analysis};
    \node[Process, below=of adjoint] (sens) {Calculate sensitivities};
    \node[Process, below=of sens] (update) {Update design variables};
    \draw[Line] (start.south) -- (init.north);
    \draw[Line] (init.south) -- (filter.north) coordinate[midway] (mid_point);
    \draw[Line] (filter.south) -- (ray tracing.north);
    \draw[Line] (ray tracing.south) -- (state.north);
    \draw[Line] (state.south) -- (obj.north);
    \draw[Line] (obj.south) -- (conv.north);
    \draw[Line] (adjoint.south) -- (sens.north);
    \draw[Line] (sens.south) -- (update.north);
    \draw[Arrow] (conv.east) -- node[above,pos=0.3] {Yes} (end.west);
    \draw[Arrow] (conv.south) -- node[right,pos=0.3] {No} (adjoint.north);
    \coordinate (bot) at ([yshift=-5mm] update.south); 
    \coordinate (left_bot) at ([xshift=-10mm] filter.west |- bot); 
    \coordinate (left_mid) at (left_bot |- mid_point); 
    \draw[Arrow] (update.south) -- (bot) -- (left_bot) -- (left_mid) -- (mid_point);
  \end{tikzpicture}
  \caption{Flowchart of the topology optimization algorithm.}
  \label{figAlgorithm}
\end{figure}

The overall TO algorithm is illustrated in the flowchart shown in \fref{figAlgorithm}.
First, the design variable vector $\bm{\psi}$ is initialized for all elements in the design domain.
The density filtering process is then performed according to \eref{eqFilter} to obtain a smooth dimensionless density distribution $\bm{\phi}$.
Based on this filtered density field, a ray-tracing procedure is carried out to evaluate the exchange factors.
Using the resulting exchange factors and density distribution, the state analysis is performed by iteratively solving the nonlinear governing equations, \eref{eqResidual}, for the dimensionless nodal temperature vector $\mathbf{T}^*$ using the Newton-Raphson method.
Finally, the objective function $J$ and the volume constraint $V^*$ are evaluated from the obtained temperature field.
\par
Subsequently, the convergence of the optimization process is evaluated.
The optimization is regarded as converged when the relative change in the objective function between two consecutive design iterations becomes smaller than a prescribed tolerance $\epsilon$, i.e.,
\begin{align}\label{eqConvergence}
  \frac{\left| J^{(k)} - J^{(k-1)} \right|}{\left| J^{(k)} \right|} \leq \epsilon,
\end{align}
where the superscript $(k)$ denotes the current design iteration.
If \eref{eqConvergence} is satisfied, the optimization process is terminated.
\par
If the convergence criterion is not satisfied, the adjoint equation, \eref{eqAdjointEq}, is solved using the current temperature field to obtain the adjoint variable vector $\bm{\lambda}$.
The total sensitivity of the objective function with respect to the design variables, $\mathrm{d}J/\mathrm{d}\psi_{\df}$, is then evaluated according to \eref{eqSensitivity}.
Based on the resulting sensitivities and the volume constraint, the design variables are updated using the Method of Moving Asymptotes (MMA)~\cite{svanbergMMA1987}.
The updated design variable vector $\bm{\psi}^{(k+1)}$ is subsequently passed to the filtering process, and the entire procedure is repeated until the convergence criterion is satisfied.

\section{Numerical examples}\label{secNumericalExample}
This section presents several numerical examples to demonstrate the effectiveness of the proposed method.
First, \sref{secOptimizationProblemSettings} describes the common numerical settings used throughout the numerical examples.
Then, \sref{subsecHeatSink2D} and \sref{subsecHeatSink3D} present the optimized designs of two- and three-dimensional radiative heat sinks, respectively, and \sref{subsecRadiationShield2D} presents the optimized designs of radiation shields.
\subsection{Numerical settings}\label{secOptimizationProblemSettings}
This subsection describes the common numerical settings and the choice of dimensionless parameters used in the following examples.
The reference temperature $T_{\rm ref}$ is defined based on the internal heat generation rate as $T_{\rm ref}=Q_{\rm in}L^2/k_{\rm max}$, which ensures that the dimensionless internal heat generation is consistently set to $Q_{\rm in}^*=1$.
Consequently, the thermal characteristics of the system are characterized by the conduction-radiation parameter $N_{\rm R}$, which represents the relative strength of radiative heat transfer compared with conductive heat transfer.
Therefore, the influence of thermal radiation is investigated by varying $N_{\rm R}$ while keeping $Q_{\rm in}^*=1$.
To avoid singularity of the stiffness matrix in the finite element analysis, the minimum dimensionless thermal conductivity of the void region is set to $k_{\min}^*=10^{-8}$.
The SIMP penalization parameters are set to $p=2$ for the thermal conductivity (\eref{eqInterpolationk}) and $q=2$ for the radiative interpolation (\eref{eqInterpolationphi}).
Because these interpolation parameters can affect the final topology, a comparison of the resulting structures is provided in \saref{appInterpolation}.
\par
To emulate a space environment, an angle-dependent radiative boundary condition is imposed on rays escaping from the design domain.
Specifically, rays emitted in directions with a positive elevation angle ($\sin\varphi>0$) escape into deep space with an ambient temperature of absolute zero ($T_{\text{env}}^*=0$), whereas rays emitted in directions with a negative elevation angle ($\sin\varphi<0$) are treated as adiabatic and reflected back into the system.
The convergence tolerance in \eref{eqConvergence} is set to $\epsilon = 1.0 \times 10^{-7}$.
These settings are common to all examples, whereas the design domain, target evaluation domain, and ray-tracing parameters are specified in each subsection.

\subsection{Optimization of two-dimensional radiative heat sink}\label{subsecHeatSink2D}
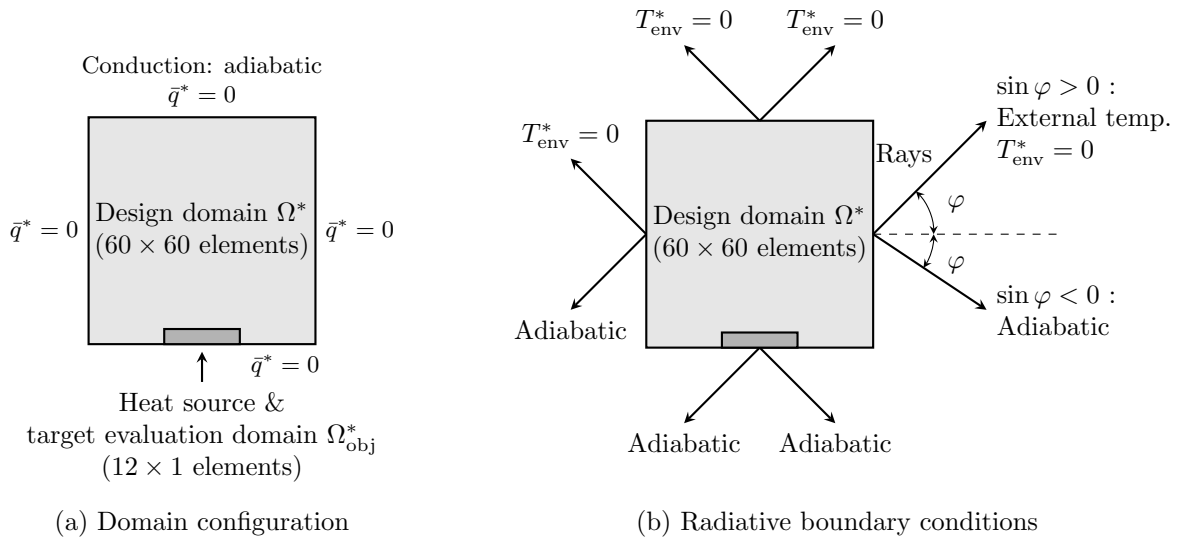
\begin{figure}[htbp]
  \centering
  \begin{subfigure}[b]{0.4\linewidth}
    \centering
    \begin{tikzpicture}[
        scale=1.0,
        font=\small,
        dim/.style={<->, >=stealth, thin},
        pointer/.style={->, >=stealth, thick, shorten >=1mm},
        ray/.style={->, >=stealth, thick, color=black}
      ]
      \path (0, 4.5) -- (0, -1.9);

      \draw[thick, color=black,fill=gray!20] (0,0) rectangle (3,3);
      \node[align=center,text=black] at (1.5, 1.5) {Design domain $\Omega^*$\\($60 \times 60$ elements)};

      \draw[thick, black, fill=gray!60] (1.0, 0) rectangle (2.0, 0.2);
      \draw[pointer, black] (1.5, -0.5) node[below, align=center] {Heat source \& \\ target evaluation domain $\Omega_{\rm obj}^*$\\($12 \times 1$ elements)} -- (1.5, 0.0);

      \node[above, font=\footnotesize, text=black,align=center] at (1.5, 3.0) {Conduction: adiabatic \\$\bar{q}^* = 0$};
      \node[left, font=\footnotesize, text=black] at (0, 1.5) {$\bar{q}^* = 0$};
      \node[right, font=\footnotesize, text=black] at (3.0, 1.5) {$\bar{q}^* = 0$};
      \node[below, font=\footnotesize, text=black] at (2.6, 0.0) {$\bar{q}^* = 0$};

    \end{tikzpicture}
    \caption{Domain configuration}
    \label{subfigRadiativeHeatSinkDomain}
  \end{subfigure}
  \hfill
  \begin{subfigure}[b]{0.55\linewidth}
    \centering
    \begin{tikzpicture}[
        scale=1.0,
        font=\small,
        dim/.style={<->, >=stealth, thin},
        pointer/.style={->, >=stealth, thick, shorten >=1mm},
        ray/.style={->, >=stealth, thick, color=black}
      ]
      \path (0, 4.5) -- (0, -1.9);

      \draw[thick, color=black,fill=gray!20] (0,0) rectangle (3,3);
      \node[align=center,text=black] at (1.5, 1.5) {Design domain $\Omega^*$\\($60 \times 60$ elements)};

      \draw[thick, black, fill=gray!60] (1.0, 0) rectangle (2.0, 0.2);

      \coordinate (OR) at (3.0, 1.5);
      \draw[dashed, thin,black] (OR) -- (5.5, 1.5);
      \draw[ray] (OR) -- (4.5, 3.0) node[midway, above left, text=black,xshift=5pt] {Rays};
      \draw[<->, >=stealth,  thin,black] (3.8, 1.5) arc[start angle=0, end angle=45, radius=0.8];
      \node[black] at (4.1, 1.9) {$\varphi$};
      \node[align=left, right,black] at (4.5, 3.0) {$\sin\varphi > 0$ :\\External temp. \\$T_{\text{env}}^* = 0$};
      \draw[ray] (OR) -- (4.5, 0.5) node[midway, below left, text=black,xshift=5pt] {};
      \draw[<->, >=stealth, thin, black] (3.8, 1.5) arc[start angle=0, end angle=-33.7, radius=0.8];
      \node[black] at (4.1, 1.1) {$\varphi$};
      \node[align=left, right,black] at (4.5, 0.5) {$\sin\varphi < 0$ :\\Adiabatic};

      \coordinate (OL) at (0.0, 1.5);
      \draw[ray] (OL) -- (-1.0, 2.5) node[midway, above right, text=black,xshift=-5pt] {};
      \node[align=right, above,black] at (-1.0, 2.5) {$T_{\text{env}}^* = 0$};
      \draw[ray] (OL) -- (-1.0, 0.5) node[midway, below right, text=black,xshift=-5pt] {};
      \node[align=right, below,black] at (-1.0, 0.5) {Adiabatic};

      \coordinate (OT) at (1.5, 3.0);
      \draw[ray] (OT) -- (0.5, 4.0);
      \draw[ray] (OT) -- (2.5, 4.0);
      \node[align=center, above, text=black] at (0.5, 4.0) {$T_{\text{env}}^* = 0$};
      \node[align=center, above, text=black] at (2.5, 4.0) {$T_{\text{env}}^* = 0$};

      \coordinate (OB) at (1.5, 0.0);
      \draw[ray] (OB) -- (0.5, -1.0);
      \draw[ray] (OB) -- (2.5, -1.0);
      \node[align=center, below, text=black] at (0.5, -1.0) {Adiabatic};
      \node[align=center, below, text=black] at (2.5, -1.0) {Adiabatic};
    \end{tikzpicture}
    \caption{Radiative boundary conditions}
    \label{subfigRadiativeHeatSinkBC}
  \end{subfigure}
  \caption{Problem settings for the two-dimensional radiative heat sink design, illustrating (a) the design domain and heat source, and (b) the angle-dependent radiative boundary conditions.}
  \label{figRadiativeHeatSink}
\end{figure}
The optimization of a two-dimensional radiative heat sink is first considered.
As illustrated in \fref{figRadiativeHeatSink}, the design domain $\Omega^*$ is a square region discretized into $60 \times 60$ finite elements.
A heat source region, which also serves as the target evaluation domain $\Omega_{\rm obj}^*$ for the objective function, is located at the bottom center of the design domain and consists of $12 \times 1$ elements.
For thermal conduction, all external boundaries of the design domain are subjected to an adiabatic condition with $\bar{q}^* = 0$.
The dimensionless filter radius is set to $R_{\min}^*=0.05$.
In the ray-tracing procedure, $N_{\rm pt}=10$ launch points per element face and $N_{\rm ang}=100$ rays per launch point are used.
The proposed TO is performed for several conduction-radiation parameters, $N_{\rm R}\in\left[10^{-8},10^{-6},10^{-4},10^{-2},1,10^2,10^4,10^6,10^8\right]$, defined in \eref{eqNR}.
The initial design variable is uniformly set to $\psi_{\te}=0.5$.
The upper limit of the dimensionless volume is set to $V_{\max}^*=0.3$.
\par
\fref{figHeatSink2D} shows the optimized structure for $N_{\rm R}=1$.
As seen in \fref{subfigHeatSink2D1}, the optimized structure consists of two thick branches extending from the heat source toward the top corners and two thin horizontal branches toward the bottom corners.
By filling the corners with solid material, this structure maximizes the projected area in all viewing directions, which dominates the radiative heat transfer.
At the same time, the branches enhance heat conduction toward the corners.
Because the top corners contribute to heat emission toward the external environment more than the bottom corners due to the angle-dependent radiative boundary conditions, thicker branches are formed toward the top corners.
\par
\fref{subfigHeatSink2D2} shows the distribution of the dimensionless volumetric net radiative heat transfer $Q_{{\rm rad}}^*$.
The radiative heat exchange is evaluated smoothly across the implicit structural surfaces.
The value of $Q_{{\rm rad}}^*$ corresponds to the surface radiative heat flux distributed over the intermediate-density layers.
The magnitude of $Q_{{\rm rad}}^*$ is therefore larger at the corners and outer boundaries of the analysis domain, where the intermediate-density layers are relatively thin.
\par
\fref{subfigHeatSink2D3} shows the temperature distribution, where the temperature gradient aligns with the direction of the branches, indicating that the structure reduces the thermal resistance between the heat source and the four corners. Owing to the minimum thermal conductivity $k_{\min}^*$, the void region also exhibits a temperature distribution intermediate between that of the solid and the external environment.
\par
\fref{subfigHeatSink2D4} shows the binarized version of the optimized structure.
To preserve the exact dimensionless volume $V^*$, the 1,080 elements with the highest dimensionless densities are assigned as solid, while the remaining elements are assigned as void.
A conduction-radiation analysis is then performed for the binarized structure, and the resulting results are presented in \frefs{subfigHeatSink2D5} and \ref{subfigHeatSink2D6}.
As shown in \fref{subfigHeatSink2D5}, the dimensionless volumetric net radiative heat transfer $Q_{\rm rad}^*$, which is distributed throughout the intermediate-density layers in the optimization model, becomes explicitly concentrated in the surface elements of the binarized model.
This localization provides a clearer representation of the radiative heat transfer distribution.
Furthermore, in the corner regions, where solid elements are directly exposed without surrounding intermediate-density elements even in the optimization model, the values of $Q_{\rm rad}^*$ are nearly identical in the two models.
\par
\fref{subfigHeatSink2D6} shows the dimensionless temperature field $T^*$ obtained for the binarized structure.
The overall temperature range and spatial distribution are consistent with those of the optimization model, particularly at the branch tips.
This agreement demonstrates that the proposed method is able to capture the overall radiative heat transfer even on surfaces represented by intermediate densities.
However, the maximum temperature near the heat source is slightly lower in the binarized model than in the optimization model.
This difference results from the penalization of thermal conductivity in the intermediate-density regions of the optimization model.
The penalization artificially reduces heat conduction near the structural boundaries, thereby increasing the thermal resistance between the heat source and the branches and, consequently, raising the temperature at the heat source.
As detailed in \saref{appInterpolation}, this penalization is necessary to prevent the optimized structures from containing extensive intermediate-density regions.
Therefore, the resulting minor difference in thermal resistance is an inherent feature of the proposed approach.
\newcommand{\HeatSinkTwoDscale}{1.05}
\newcommand{\HeatSinkTwoDwidth}{0.3\linewidth}
\begin{figure}[htbp]
  \centering
  \begin{subfigure}[t]{\HeatSinkTwoDwidth}
    \centering
    \includegraphics[scale=\HeatSinkTwoDscale,page=1]{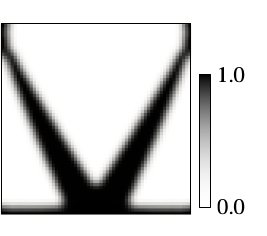}
    \caption{Dimensionless density}
    \label{subfigHeatSink2D1}
  \end{subfigure}
  \hfill
  \begin{subfigure}[t]{\HeatSinkTwoDwidth}
    \centering
    \includegraphics[scale=\HeatSinkTwoDscale,page=2]{z401_HeatSink2D_NR1.pdf}
    \caption{Dimensionless volumetric net radiative heat transfer $Q_{{\rm rad}}^*$}
    \label{subfigHeatSink2D2}
  \end{subfigure}
  \hfill
  \begin{subfigure}[t]{\HeatSinkTwoDwidth}
    \centering
    \includegraphics[scale=\HeatSinkTwoDscale,page=3]{z401_HeatSink2D_NR1.pdf}
    \caption{Dimensionless temperature $T^*$}
    \label{subfigHeatSink2D3}
  \end{subfigure}
  \\\vspace{2em}
  \centering
  \begin{subfigure}[t]{\HeatSinkTwoDwidth}
    \centering
    \includegraphics[scale=\HeatSinkTwoDscale,page=4]{z401_HeatSink2D_NR1.pdf}
    \caption{Binarized structure }
    \label{subfigHeatSink2D4}
  \end{subfigure}
  \hfill
  \begin{subfigure}[t]{\HeatSinkTwoDwidth}
    \centering
    \includegraphics[scale=\HeatSinkTwoDscale,page=5]{z401_HeatSink2D_NR1.pdf}
    \caption{Dimensionless volumetric net radiative heat transfer $Q_{{\rm rad}}^*$ calculated for the binarized structure}
    \label{subfigHeatSink2D5}
  \end{subfigure}
  \hfill
  \begin{subfigure}[t]{\HeatSinkTwoDwidth}
    \centering
    \includegraphics[scale=\HeatSinkTwoDscale,page=6]{z401_HeatSink2D_NR1.pdf}
    \caption{Dimensionless temperature $T^*$ calculated for the binarized structure}
    \label{subfigHeatSink2D6}
  \end{subfigure}
  \caption{Optimized two-dimensional radiative heat sink structure for $N_{\rm R}=1$.}
  \label{figHeatSink2D}
\end{figure}
\begin{figure}[htbp]
  \begin{center}
    \includegraphics[scale=1.0]{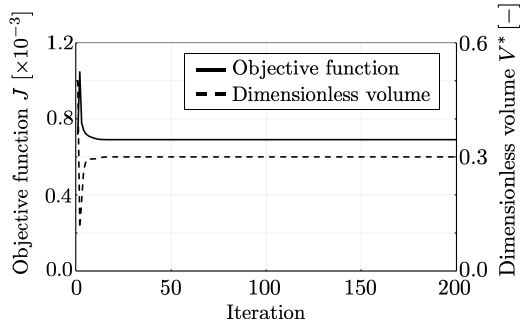}
  \end{center}
  \caption{Optimization history of the objective function and volume constraint of the case in \fref{figHeatSink2D}.}
  \label{figConvergence}
\end{figure}
\par
\fref{figConvergence} plots the optimization histories of the objective function $J$ and dimensionless volume $V^*$ during the optimization.
The objective function smoothly converges to $J=6.9041\times 10^{-4}$ while the volume constraint remains active throughout the optimization process.
In addition, \fref{figHeatSink2DNR} shows the optimized structures for various values of the conduction-radiation parameter $N_{\rm R}$.
As defined in \eref{eqNR}, a larger $N_{\rm R}$ corresponds to a lower thermal conductivity, and  vice versa.
As seen in \frefs{subfigHeatSink2DNR4}-\ref{subfigHeatSink2DNR8}, the optimized structures do not reach the top corners when $N_{\rm R}$ is large.
Under such low-conductivity conditions, transferring heat efficiently from the heat source to the top corners is difficult, and radiative heat transfer from the high-temperature region near the heat source is used instead.
Conversely, for small values of $N_{\rm R}$, the branches of the optimized structures exhibit irregular bending, as shown in \frefs{subfigHeatSink2DNRm8}, \ref{subfigHeatSink2DNRm6}.
In these cases, the thermal conductivity is sufficiently large, so that the temperature distribution becomes nearly uniform.
The conduction term in the sensitivity therefore becomes small, and the optimized structure is strongly affected by the angular discretization errors; refer to \saref{appdiscretizationError} for details.
\newcommand{\HeatSinkTwoDNRscale}{0.5}
\newcommand{\HeatSinkTwoDNRwidth}{0.23\linewidth}
\begin{figure}[htbp]
  \centering
  \begin{subfigure}[t]{\HeatSinkTwoDNRwidth}
    \centering
    \includegraphics[scale=\HeatSinkTwoDNRscale,page=1]{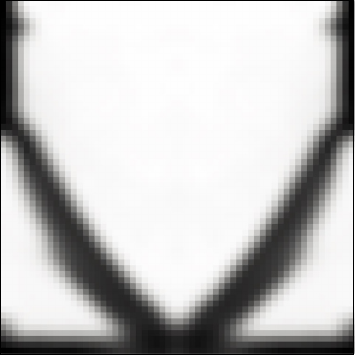}
    \caption{$N_{\rm R}=10^{-8}$}
    \label{subfigHeatSink2DNRm8}
  \end{subfigure}
  \hspace{3em}
  \begin{subfigure}[t]{\HeatSinkTwoDNRwidth}
    \centering
    \includegraphics[scale=\HeatSinkTwoDNRscale,page=2]{z401_HeatSink2D_NR.pdf}
    \caption{$N_{\rm R}=10^{-6}$}
    \label{subfigHeatSink2DNRm6}
  \end{subfigure}
  \hspace{3em}
  \begin{subfigure}[t]{\HeatSinkTwoDNRwidth}
    \centering
    \includegraphics[scale=\HeatSinkTwoDNRscale,page=3]{z401_HeatSink2D_NR.pdf}
    \caption{$N_{\rm R}=10^{-4}$}
    \label{subfigHeatSink2DNRm4}
  \end{subfigure}
  \\\vspace{1em}
  \begin{subfigure}[t]{\HeatSinkTwoDNRwidth}
    \centering
    \includegraphics[scale=\HeatSinkTwoDNRscale,page=4]{z401_HeatSink2D_NR.pdf}
    \caption{$N_{\rm R}=10^{-2}$}
    \label{subfigHeatSink2DNRm2}
  \end{subfigure}
  \hspace{3em}
  \begin{subfigure}[t]{\HeatSinkTwoDNRwidth}
    \centering
    \includegraphics[scale=\HeatSinkTwoDNRscale,page=5]{z401_HeatSink2D_NR.pdf}
    \caption{$N_{\rm R}=1$}
    \label{subfigHeatSink2DNR0}
  \end{subfigure}
  \hspace{3em}
  \begin{subfigure}[t]{\HeatSinkTwoDNRwidth}
    \centering
    \includegraphics[scale=\HeatSinkTwoDNRscale,page=6]{z401_HeatSink2D_NR.pdf}
    \caption{$N_{\rm R}=10^{2}$}
    \label{subfigHeatSink2DNR2}
  \end{subfigure}
  \\\vspace{1em}
  \begin{subfigure}[t]{\HeatSinkTwoDNRwidth}
    \centering
    \includegraphics[scale=\HeatSinkTwoDNRscale,page=7]{z401_HeatSink2D_NR.pdf}
    \caption{$N_{\rm R}=10^{4}$}
    \label{subfigHeatSink2DNR4}
  \end{subfigure}
  \hspace{3em}
  \begin{subfigure}[t]{\HeatSinkTwoDNRwidth}
    \centering
    \includegraphics[scale=\HeatSinkTwoDNRscale,page=8]{z401_HeatSink2D_NR.pdf}
    \caption{$N_{\rm R}=10^{6}$}
    \label{subfigHeatSink2DNR6}
  \end{subfigure}
  \hspace{3em}
  \begin{subfigure}[t]{\HeatSinkTwoDNRwidth}
    \centering
    \includegraphics[scale=\HeatSinkTwoDNRscale,page=9]{z401_HeatSink2D_NR.pdf}
    \caption{$N_{\rm R}=10^{8}$}
    \label{subfigHeatSink2DNR8}
  \end{subfigure}
  \caption{Optimized two-dimensional radiative heat sink structures for various values of $N_{\rm R}$.}
  \label{figHeatSink2DNR}
\end{figure}

\subsection{Optimization of three-dimensional radiative heat sink}\label{subsecHeatSink3D}
\begin{figure}[htbp]
  \centering
  \begin{tikzpicture}[
      scale=1.5,
      font=\small,
      x={(1cm,0cm)}, y={(0cm,1cm)}, z={(0.4cm,0.3cm)},
      pointer/.style={->, >=stealth, thick, shorten >=1mm, black},
      ray/.style={->, >=stealth, thick, color=black}
    ]
    \draw[fill=gray!15, draw=none] (0,0,0) -- (2.5,0,0) -- (2.5,0,2.5) -- (0,0,2.5) -- cycle;
    \draw[fill=gray!5, draw=none] (0,0,0) -- (0,0,2.5) -- (0,2.5,2.5) -- (0,2.5,0) -- cycle;
    \draw[fill=gray!10, draw=none] (0,0,2.5) -- (2.5,0,2.5) -- (2.5,2.5,2.5) -- (0,2.5,2.5) -- cycle;
    \draw[dashed, thin, black] (0,0,0) -- (0,0,2.5);
    \draw[dashed, thin, black] (0,0,2.5) -- (2.5,0,2.5);
    \draw[dashed, thin, black] (0,0,2.5) -- (0,2.5,2.5);
    \draw[thick, black, fill=gray!60,dashed, dash pattern=on 2pt off 2pt,preaction={draw=gray!60,thick}] (0.8,0,0.8) -- (1.7,0,0.8) -- (1.7,0.15,0.8) -- (0.8,0.15,0.8) -- cycle;
    \draw[thick, black, fill=gray!50,dashed, dash pattern=on 2pt off 2pt,preaction={draw=gray!50,thick}] (1.7,0,0.8) -- (1.7,0,1.7) -- (1.7,0.15,1.7) -- (1.7,0.15,0.8) -- cycle;
    \draw[thick, black, fill=gray!40,dashed, dash pattern=on 2pt off 2pt,preaction={draw=gray!40,thick}] (0.8,0.15,0.8) -- (1.7,0.15,0.8) -- (1.7,0.15,1.7) -- (0.8,0.15,1.7) -- cycle;
    \draw[thick, black] (0,0,0) -- (2.5,0,0) -- (2.5,2.5,0) -- (0,2.5,0) -- cycle;
    \draw[thick, black] (2.5,0,0) -- (2.5,0,2.5) -- (2.5,2.5,2.5) -- (2.5,2.5,0);
    \draw[thick, black] (0,2.5,0) -- (2.5,2.5,0) -- (2.5,2.5,2.5) -- (0,2.5,2.5) -- cycle;
    \node[align=center, text=black,fill=white] at (1.25, 1.5, 0) {Design domain $\Omega^*$\\($30 \times 30 \times 30$ elements)};
    \draw[pointer] (-0.7, 0.75, 1.25) node[left, align=center, text=black] {Heat source \& \\ target evaluation domain $\Omega_{\rm obj}^*$\\($6 \times 6 \times 1$ elements)} -- (1.25, 0.075, 1.25);
    \coordinate (O) at (2.5, 1.25, 1.25);
    \draw[thick, black, fill=gray!0,dashed] (2.5, 1.0, 1.0) -- (2.5, 1.5, 1.0) -- (2.5, 1.5, 1.5) -- (2.5, 1.0, 1.5) -- cycle;
    \draw[dashed, thin, black] (O) -- (4.2, 1.25, 1.25);
    \draw[ray] (O) -- (3.8, 2.3, 1.25) node[midway, above left, text=black, xshift=20,yshift=12] {Rays};
    \draw[<->, >=stealth, thin, black] (3.2, 1.25, 1.25) arc[start angle=0, end angle=38.9, radius=0.7];
    \node[black] at (3.35, 1.5, 1.25) {$\varphi$};
    \node[align=left, right, black] at (3.8, 2.3, 1.25) {$\sin\varphi > 0$ :\\External temp. \\$T_{\text{env}}^* = 0$};
    \draw[ray] (O) -- (3.8, 0.2, 1.25) node[midway, below left, text=black, xshift=20,yshift=-12] {};
    \draw[<->, >=stealth, thin, black] (3.2, 1.25, 1.25) arc[start angle=0, end angle=-38.9, radius=0.7];
    \node[black] at (3.35, 1.0, 1.25) {$\varphi$};
    \node[align=left, right, black] at (3.8, 0.2, 1.25) {$\sin\varphi < 0$ :\\Adiabatic};

    \node[align=center, text=black] at (-1.0, 2.7, 1.5) {Conduction: adiabatic \\ $\bar{q}^* = 0$};

    \coordinate (OT) at (1.25, 2.5, 1.25);
    \draw[thick, black, fill=gray!0, dashed] (1.0, 2.5, 1.0) -- (1.5, 2.5, 1.0) -- (1.5, 2.5, 1.5) -- (1.0, 2.5, 1.5) -- cycle;
    \draw[ray] (OT) -- (1.8, 3.3, 1.25) node[above, align=left, text=black] {$T_{\text{env}}^* = 0$};

    \coordinate (OB) at (2.1, 0, 1.25);
    \coordinate (OB_mid) at (2.22, -0.21, 1.415); 
    \coordinate (OB_end) at (2.5, -0.7, 1.8);
    \draw[thick, black, fill=gray!0, dashed] (1.85, 0, 1.0) -- (2.35, 0, 1.0) -- (2.35, 0, 1.5) -- (1.85, 0, 1.5) -- cycle;

    \draw[thick, black, dashed] (OB) -- (OB_mid);
    \draw[ray] (OB_mid) -- (OB_end) node[right, align=left, text=black] {Adiabatic};
    \draw[ thick, black] (2.5,0,0) -- (2.5,2.5,0);

  \end{tikzpicture}
  \caption{Problem settings for the three-dimensional radiative heat sink design, illustrating the design domain, the internal heat source, the target evaluation domain, and the angle-dependent radiative boundary conditions.}
  \label{figRadiativeHeatSink3D}
\end{figure}
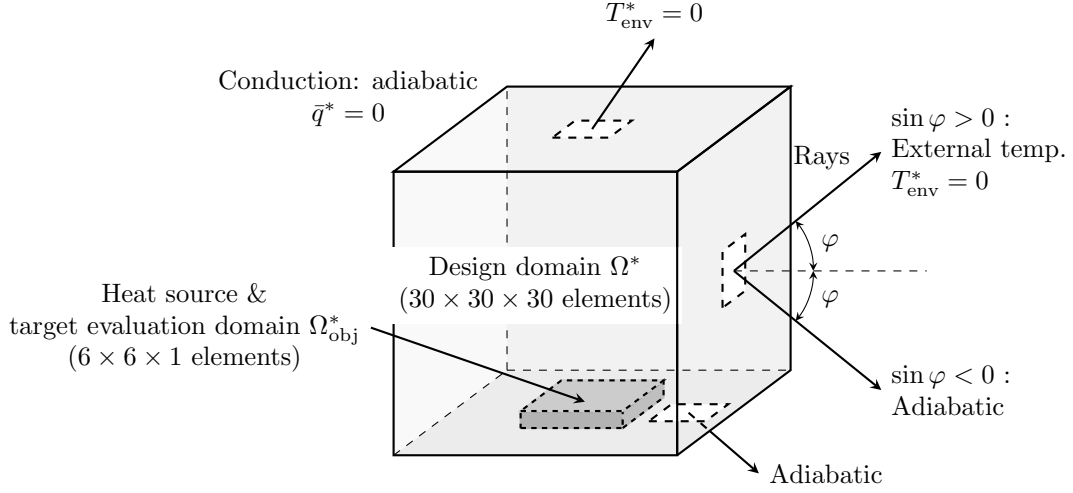
To demonstrate the applicability of the proposed method to three-dimensional problems, TO is performed for a 3D extension of the preceding 2D heat sink problem. The formulation is common to the two- and three-dimensional cases, except for the solid-angle discretization used in the view-factor evaluation, as described in \sref{secViewFactor}.
As shown in \fref{figRadiativeHeatSink3D}, the cubic design domain $\Omega^*$ is discretized into $30\times30\times30$ finite elements.
The heat source, which also serves as the target domain $\Omega_{\rm obj}^*$, is located at the bottom center and consists of $6\times6\times1$ elements.
The radiative boundary condition is the same as that in \sref{subsecHeatSink2D}.
The filter radius is set to $R_{\min}^*=0.05$.
For ray tracing, one launch point is placed at the center of each element face, and $N_{\rm ang}=240$ rays are used per launch point by discretizing the solid angle into 20 azimuthal and 12 zenith angles.
The conduction-radiation parameter is set to $N_{\rm R}=1$.
The initial design variable is $\psi_{\te}=0.5$ uniformly, and the volume constraint is $V_{\max}^*=0.3$.
\par
\newcommand{\HeatSinkThreeDscale}{1}
\begin{figure}[htbp]
  \centering
  \begin{subfigure}[t]{0.45\linewidth}
    \centering
    \includegraphics[scale=\HeatSinkThreeDscale,page=1]{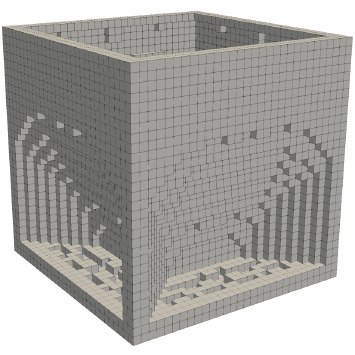}
    \caption{Optimized structure}
    \label{subfigHeatSink3D1}
  \end{subfigure}
  \hspace{0.05\linewidth}
  \begin{subfigure}[t]{0.45\linewidth}
    \centering
    \includegraphics[scale=\HeatSinkThreeDscale,page=2]{z402_HeatSink3D.pdf}
    \caption{Optimized structure clipped at the central cross-section}
    \label{subfigHeatSink3D2}
  \end{subfigure}
  \caption{Optimized three-dimensional radiative heat sink structure for $N_{\rm R}=1$.}
  \label{figHeatSink3D}
\end{figure}
\fref{figHeatSink3D} shows the optimized structure for the 3D radiative heat sink problem.
As shown in \fref{subfigHeatSink3D1}, all the twelve edges of the design domain are filled with solid material.
Consistent with the 2D cases, this structure maximizes the projected area in all viewing directions.
As seen in \fref{subfigHeatSink3D2}, the cross-sectional structure is similar to the 2D results: thick branches extend from the heat source toward the edges of the design domain.
This result demonstrates that the optimized 3D radiative heat sink tends to exhibit the same functionality as observed in 2D results.
\par
The computational cost of optimizing the two- and three-dimensional radiative heat sinks is summarized in \tref{tabComputationTime}.
Although the number of traced rays in the three-dimensional case is about three times that of the two-dimensional case, the computation time increases by a larger factor. This is mainly because the larger number of elements raises the cost of solving the linear system, whose tangent matrix becomes dense due to the radiative interactions, as noted in \sref{secDiscretization}.
\begin{table}[htbp]
  \centering
  \caption{Computation time per 100 design iterations. (Environment: Apple M3 Ultra (24 performance cores), 512~GB RAM)}
  \label{tabComputationTime}
  \begin{tabular}{lccc}
    \hline
    Problem                                       & Number of elements & Number of rays     & Time per 100 iterations [min] \\
    \hline
    2D radiative heat sink ($60\times60$)         & $3,600$            & $1.44\times10^{7}$ & 70                            \\
    3D radiative heat sink ($30\times30\times30$) & $27,000$           & $3.89\times10^{7}$ & 400                           \\
    \hline
  \end{tabular}
\end{table}

\subsection{Optimization of radiation shield}\label{subsecRadiationShield2D}
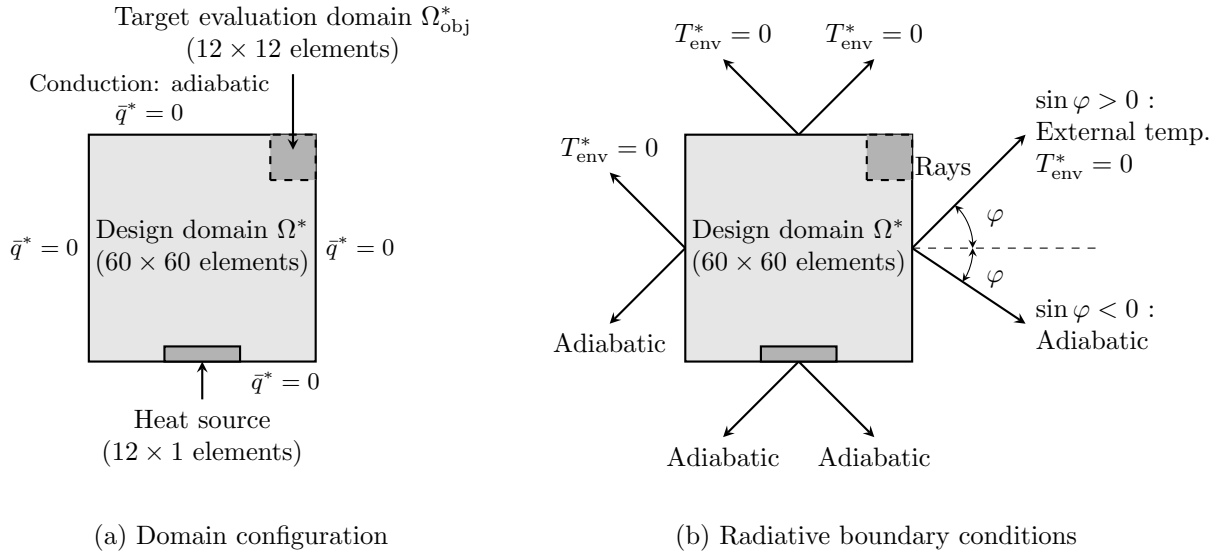
\begin{figure}[htbp]
  \centering
  \begin{subfigure}[b]{0.4\linewidth}
    \centering
    \begin{tikzpicture}[
        scale=1.0,
        font=\small,
        dim/.style={<->, >=stealth, thin},
        pointer/.style={->, >=stealth, thick, shorten >=1mm},
        ray/.style={->, >=stealth, thick, color=black}
      ]
      \path (0, 4.5) -- (0, -1.9);

      \draw[thick, color=black,fill=gray!20] (0,0) rectangle (3,3);
      \node[align=center,text=black] at (1.5, 1.5) {Design domain $\Omega^*$\\($60 \times 60$ elements)};

      \draw[thick, black, fill=gray!60] (1.0, 0) rectangle (2.0, 0.2);
      \draw[pointer, black] (1.5, -0.5) node[below, align=center] {Heat source\\($12 \times 1$ elements)} -- (1.5, 0.1);

      \draw[thick, dashed, black, fill=gray!60] (2.4, 2.4) rectangle (3.0,3.0);
      \draw[pointer, black] (2.7, 3.8) node[above, align=center] {Target evaluation domain $\Omega_{\rm obj}^*$\\($12 \times 12$ elements)} -- (2.7, 2.7);

      \node[above, font=\footnotesize, text=black,align=center] at (0.8, 3.0) {Conduction: adiabatic \\$\bar{q}^* = 0$};
      \node[left, font=\footnotesize, text=black] at (0, 1.5) {$\bar{q}^* = 0$};
      \node[right, font=\footnotesize, text=black] at (3.0, 1.5) {$\bar{q}^* = 0$};
      \node[below, font=\footnotesize, text=black] at (2.6, 0.0) {$\bar{q}^* = 0$};

    \end{tikzpicture}
    \caption{Domain configuration}
    \label{subfigRadiationShieldDomain}
  \end{subfigure}
  \hfill
  \begin{subfigure}[b]{0.55\linewidth}
    \centering
    \begin{tikzpicture}[
        scale=1.0,
        font=\small,
        dim/.style={<->, >=stealth, thin},
        pointer/.style={->, >=stealth, thick, shorten >=1mm},
        ray/.style={->, >=stealth, thick, color=black}
      ]
      \path (0, 4.5) -- (0, -1.9);

      \draw[thick, color=black,fill=gray!20] (0,0) rectangle (3,3);
      \node[align=center,text=black] at (1.5, 1.5) {Design domain $\Omega^*$\\($60 \times 60$ elements)};
      \draw[thick, black, fill=gray!60] (1.0, 0) rectangle (2.0, 0.2);

      \draw[thick, dashed, black, fill=gray!60] (2.4, 2.4) rectangle (3.0,3.0);

      \coordinate (OR) at (3.0, 1.5);
      \draw[dashed, thin,black] (OR) -- (5.5, 1.5);
      \draw[ray] (OR) -- (4.5, 3.0) node[midway, above left, text=black,xshift=5pt] {Rays};
      \draw[<->, >=stealth,  thin,black] (3.8, 1.5) arc[start angle=0, end angle=45, radius=0.8];
      \node[black] at (4.1, 1.9) {$\varphi$};
      \node[align=left, right,black] at (4.5, 3.0) {$\sin\varphi > 0$ :\\External temp. \\$T_{\text{env}}^* = 0$};
      \draw[ray] (OR) -- (4.5, 0.5) node[midway, below left, text=black,xshift=5pt] {};
      \draw[<->, >=stealth, thin, black] (3.8, 1.5) arc[start angle=0, end angle=-33.7, radius=0.8];
      \node[black] at (4.1, 1.1) {$\varphi$};
      \node[align=left, right,black] at (4.5, 0.5) {$\sin\varphi < 0$ :\\Adiabatic};

      \coordinate (OL) at (0.0, 1.5);
      \draw[ray] (OL) -- (-1.0, 2.5) node[midway, above right, text=black,xshift=-5pt] {};
      \node[align=right, above,black] at (-1.0, 2.5) {$T_{\text{env}}^* = 0$};
      \draw[ray] (OL) -- (-1.0, 0.5) node[midway, below right, text=black,xshift=-5pt] {};
      \node[align=right, below,black] at (-1.0, 0.5) {Adiabatic};

      \coordinate (OT) at (1.5, 3.0);
      \draw[ray] (OT) -- (0.5, 4.0);
      \draw[ray] (OT) -- (2.5, 4.0);
      \node[align=center, above, text=black] at (0.5, 4.0) {$T_{\text{env}}^* = 0$};
      \node[align=center, above, text=black] at (2.5, 4.0) {$T_{\text{env}}^* = 0$};

      \coordinate (OB) at (1.5, 0.0);
      \draw[ray] (OB) -- (0.5, -1.0);
      \draw[ray] (OB) -- (2.5, -1.0);
      \node[align=center, below, text=black] at (0.5, -1.0) {Adiabatic};
      \node[align=center, below, text=black] at (2.5, -1.0) {Adiabatic};
    \end{tikzpicture}
    \caption{Radiative boundary conditions}
    \label{subfigRadiationShieldBC}
  \end{subfigure}
  \caption{Problem settings for the radiation shield design, illustrating (a) the design domain, internal heat source, and target evaluation domain, and (b) the angle-dependent radiative boundary conditions.}
  \label{figRadiationShield}
\end{figure}
The proposed method is subsequently applied to the design of a radiation shield.
As illustrated in \fref{figRadiationShield}, this problem aims to suppress heat transfer to a specific target evaluation domain located near a heat source.
The design domain $\Omega^*$ is a square region discretized into $60 \times 60$ finite elements.
A heat source region is located at the bottom center of the design domain and comprises $12 \times 1$ elements.
The target evaluation domain $\Omega_{\rm obj}^*$ consisting of $12\times12$ elements is located at the upper-right region.
The radiative boundary condition is the same as in the 2D heat sink design in \sref{subsecHeatSink2D}.
To obtain finer structures, the dimensionless filter radius is set to $R_{\min}^* = 0.02$.
In the ray-tracing procedure, $N_{\rm pt} = 10$ launch points per element face and $N_{\rm ang} = 100$ rays per launch point are used.
The optimization is performed for the conduction-radiation parameters $N_{\rm R}\in\left[1,10^2,10^4\right]$.
In this example, the initial design variable is uniformly set to $\psi_{\te}=1.0$.
No volume constraint is imposed in this radiation shield design problem, since a fully solid structure cannot be optimal for shielding.
\newcommand{\ShieldTwoDscale}{0.8}
\begin{figure}[htbp]
  \centering
  \begin{subfigure}[t]{0.3\linewidth}
    \centering
    \includegraphics[scale=\ShieldTwoDscale,page=2]{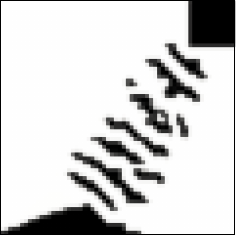}
    \caption{ $N_{\rm R}=1$}
    \label{subfigShield2D1}
  \end{subfigure}
  \hspace{0.03\linewidth}
  \begin{subfigure}[t]{0.3\linewidth}
    \centering
    \includegraphics[scale=\ShieldTwoDscale,page=3]{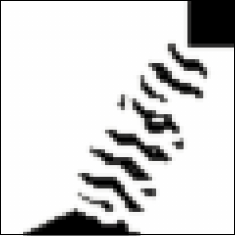}
    \caption{$N_{\rm R}=10^{2}$}
    \label{subfigShield2D2}
  \end{subfigure}
  \hspace{0.03\linewidth}
  \begin{subfigure}[t]{0.3\linewidth}
    \centering
    \includegraphics[scale=\ShieldTwoDscale,page=4]{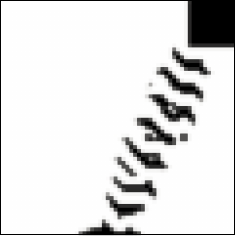}
    \caption{$N_{\rm R}=10^{4}$}
    \label{subfigShield2D3}
  \end{subfigure}
  \caption{Optimized radiation shield structures.}
  \label{figShield2D}
\end{figure}
\newcommand{\ShieldTwoDTscale}{1.05}
\newcommand{\ShieldTwoDTwidth}{0.3\linewidth}
\begin{figure}[htbp]
  \centering
  \begin{subfigure}[t]{\ShieldTwoDTwidth}
    \centering
    \includegraphics[scale=\ShieldTwoDTscale,page=1]{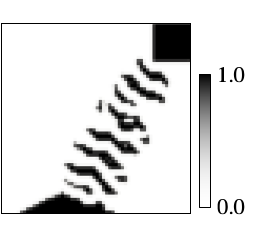}
    \caption{Optimized radiation shield structure}
    \label{subfigShield2DT1}
  \end{subfigure}
  \hspace{0.03\linewidth}
  \begin{subfigure}[t]{\ShieldTwoDTwidth}
    \centering
    \includegraphics[scale=\ShieldTwoDTscale,page=2]{z403_Shield2DTempQ.pdf}
    \caption{Dimensionless volumetric net radiative heat transfer $Q_{{\rm rad}}^*$}
    \label{subfigShield2DT2}
  \end{subfigure}
  \hspace{0.03\linewidth}
  \begin{subfigure}[t]{\ShieldTwoDTwidth}
    \centering
    \includegraphics[scale=\ShieldTwoDTscale,page=3]{z403_Shield2DTempQ.pdf}
    \caption{Dimensionless temperature $T^*$}
    \label{subfigShield2DT3}
  \end{subfigure}
  \caption{Optimized radiation shield, dimensionless volumetric net radiative heat transfer $Q_{{\rm rad}}^*$, and dimensionless temperature $T^*$  of the case in \fref{subfigShield2D2} ($N_{\rm R}=10^2$).}
  \label{figShield2DT}
\end{figure}
\par
\fref{figShield2D} shows the optimized radiation shield structures for each conduction-radiation parameter $N_{\rm R}$.
In all the optimized structures, layered structures form between the heat source and the target evaluation domain.
These layered configurations increase the overall thermal resistance by forcing heat to cross multiple gaps via thermal radiation.
Such features resemble multilayer insulation (MLI), which is widely used in spacecraft to protect electronic devices from excessive radiative heat transfer \cite{bapat1990experimental}.
Although practical MLI systems typically employ highly reflective materials, which deviates from the black-body assumption used in this study, the fundamental mechanism of radiative shielding is reproduced by the proposed method.
\par
Also, a difference in structural formation is observed in the lower-left region.
In \frefs{subfigShield2D1} and \ref{subfigShield2D2}, where $N_{\rm R}$ is smaller and radiation is less dominant, solid regions extending from the heat source toward the lower-left corner are observed.
These structures enhance the radiative heat transfer toward the external environment through the increased projected area.
By lowering the temperature through the enhanced radiative heat transfer, the structure also reduces the heat transfer toward the target evaluation domain.
In contrast, such a structure is not observed in \fref{subfigShield2D3}, where $N_{\rm R}$ is larger and radiation is more dominant.
Because the heat is efficiently radiated toward the external environment in this case, the optimized structure adopts a narrow geometry, thereby reducing the view factor between the heat source and the target evaluation domain.
These differences in the optimized structure with respect to $N_{\rm R}$ confirm that the optimal design strategy for radiative devices depends on the temperature, thermal conductivity, and length scale that constitute $N_{\rm R}$ in \eref{eqNR}.
\par
\frefs{subfigShield2DT2} and \ref{subfigShield2DT3} show the dimensionless volumetric net radiative heat transfer $Q_{\rm rad}^*$ and the dimensionless temperature $T^*$ for the optimized radiation shield structure with $N_{\rm R}=10^2$, namely, the case in \fref{subfigShield2D2}.
As shown in \fref{subfigShield2DT2}, each structural layer absorbs radiative energy on its lower surface, where $Q_{\rm rad}^* < 0$, and emits radiative energy from its upper surface, where $Q_{\rm rad}^* > 0$.
The absolute magnitude of the net radiative heat transfer progressively decreases in the layers located farther from the heat source.
Accordingly, the temperature distribution in \fref{subfigShield2DT3} exhibits step-like decreases across successive layers with increasing distance from the heat source.
These results demonstrate that the optimized multilayer structure effectively suppresses radiative heat transfer toward the target domain.
\par
It should be noted that this example exhibits numerical instabilities. As shown in \fref{subfigShield2D1}, the layered structures have jagged boundaries, and some layers contain small internal voids.
These features are considered nonphysical numerical artifacts.
The instability is attributed to competition between the sensitivities associated with suppressing heat conduction and forming an effective radiation shield.
Mitigating these instabilities remains an unresolved issue and should be addressed in future work.

\section{Conclusion}\label{secConclusion}
This study proposed a density-based TO method for conduction-radiation heat transfer problems by integrating a ray-tracing approach into the optimization framework. The contributions of this study are summarized as follows:
\begin{itemize}
  \item The conduction-radiation analysis model was formulated within the FEM framework based on the zonal method, in which each finite element is treated as an isothermal zone, and the radiative heat transfer between elements is evaluated through exchange factors.
  \item The differentiable ray-tracing formulation was developed to evaluate exchange factors in spatially varying density fields, enabling the multidirectional mutual radiation in intermediate-density regions and the analytical derivation of design sensitivities.
  \item The proposed method was applied to the design of radiative heat sinks.
        The optimized designs maximized the projected area to enhance radiative dissipation, and the dependence of the optimized structure on the conduction-radiation parameter $N_{\rm R}$ was clearly observed.
  \item The proposed method was also applied to the design of radiation shields.
        The optimized designs formed layered structures resembling multilayer insulation, demonstrating that the proposed method reproduces the fundamental mechanism of radiative shielding.
\end{itemize}
\par
A few limitations remain to be addressed in future work.
First, in the radiation shield problem, numerical instability was observed in the layered structures, arising from the conflicting sensitivities between heat conduction and radiation. Resolving this instability would improve the reliability of the optimized designs.
Second, this study assumed that the solid behaves as a black body and that the void is a vacuum.
Extending the formulation to account for gray or wavelength-dependent radiative properties would broaden the applicability to the optimal design considering practical materials.

\section*{Declaration of competing interest}
The authors declare that they have no known competing financial interests or personal relationships that could have appeared to influence the work reported in this paper.
\section*{Data availability}
Data will be made available on request.
\section*{Acknowledgements}
This work was supported by JSPS KAKENHI Grant Numbers 23K26017, 25K17528, and 26K00854.
\section*{Declaration of generative AI and AI-assisted technologies in the writing process}
The authors used ChatGPT (OpenAI) and Gemini (Google) to improve the grammar and readability of the manuscript, but all scientific content was developed by the authors.

\appendix

\section{Verification of the analysis model}\label{appVerificationSimulation}
This appendix verifies the validity of the proposed method through three benchmark problems.
\sref{subsecBenchmark1} examines the radiative energy conservation and the treatment of void and intermediate-density regions using a single radiating body.
\sref{subsecBenchmark2} verifies the evaluation of mutual radiation and shadowing effects using parallel plate configurations.
\sref{subsecBenchmark3} confirms the capability of the proposed method to solve coupled conduction-radiation problems through a one-dimensional radiating fin.
The computational results are compared with analytical or reference solutions.
\subsection{Thermal radiation from a single body}\label{subsecBenchmark1}
Radiative energy conservation for energy escaping to the external environment, as well as the effects of void regions ($\phi=0$) and intermediate density regions ($0<\phi<1$), are examined for the proposed ray-tracing model.
Consider a square solid body with a dimensionless side length of $L_0^* = 1$, subjected to a uniform internal heat generation rate $Q_{\rm in}^*$ and placed in an environment with an absolute zero ambient temperature ($T_{\text{env}}^* = 0$).
By setting the conduction-radiation parameter $N_{\rm R}$ to a sufficiently small value, heat conduction becomes dominant over radiative heat transfer; refer to \erefs{eqnondimGov} and \eqref{eqQradi}.
Under this condition, the temperature within the solid region can be regarded as uniform, and the steady-state equilibrium temperature can be analytically derived by balancing the internal heat generation with the radiative heat emission from the geometric surface.
The resulting analytical equilibrium temperature is given by
\begin{align}\label{eqtempana}
  T_{\text{analytical}}^* = \left(\frac{Q_{\rm in}^* V^*}{N_{\rm R} A^*}\right)^{1/4},
\end{align}
where $A^*$ and $V^*$ denote the dimensionless surface area and volume of the body, respectively.
For $N_{\rm R} =  10^{-4}$ and $Q_{\rm in}^* = 1.0 $, this yields $T_{\text{analytical}}^* = 7.07107$.
In all configurations, $N_{\rm pt} = 10$ launch points per element face and $N_{\rm ang} = 100$ rays per launch point are used.
\par
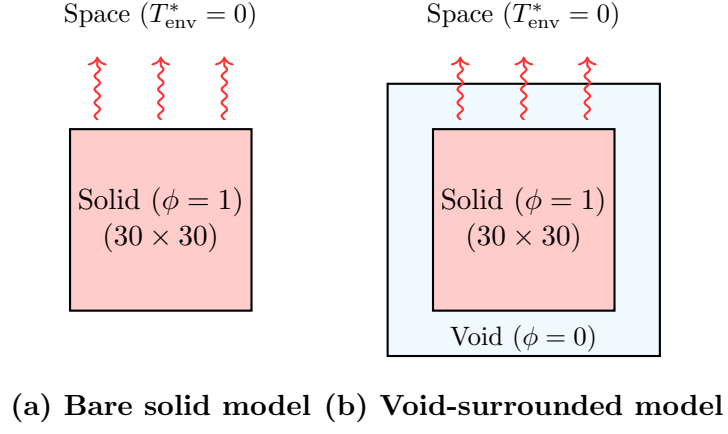
\begin{figure}[htbp]
  \centering
  \begin{tikzpicture}[scale=1.2,color=black]
    \begin{scope}[shift={(0,0)}]
      \draw[thick, fill=red!20] (0,0) rectangle (2,2);
      \node[align=center] at (1, 1) {Solid $(\phi=1)$\\($30 \times 30$)};
      \node[above, font=\small, text=black] at (1, 3) {Space ($T^*_{\rm env}=0$)};
      \foreach \x in {0.3, 1, 1.7} {
          \draw[->, red!80, thick, decorate, decoration={snake, amplitude=1.0, segment length=5, post length=1mm}] (\x, 2.1) -- (\x, 2.8);
        }
      \node[below, font=\bfseries] at (1, -0.8) {(a) Bare solid model};
    \end{scope}
    \begin{scope}[shift={(4,0)}]
      \draw[thick, fill=cyan!5] (-0.5,-0.5) rectangle (2.5,2.5);
      \node[font=\small] at (1, -0.3) {Void $(\phi=0)$};
      \draw[thick, fill=red!20] (0,0) rectangle (2,2);
      \node[align=center] at (1, 1) {Solid $(\phi=1)$\\($30 \times 30$)};
      \node[above, font=\small, text=black] at (1, 3) {Space ($T^*_{\rm env}=0$)};
      \foreach \x in {0.3, 1, 1.7} {
          \draw[->, red!80, thick, decorate, decoration={snake, amplitude=1.0, segment length=5, post length=1mm}] (\x, 2.1) -- (\x, 2.8);
        }
      \node[below, font=\bfseries] at (1, -0.8) {(b) Void-surrounded model};
    \end{scope}
  \end{tikzpicture}
  \caption{Comparison of geometric settings for the void verification models.}
  \label{figVerificationVoid}
\end{figure}

First, to verify both radiative energy conservation and the perfect transmissivity of void regions, two discrete configurations are considered, as illustrated in \fref{figVerificationVoid}.
The first is a bare solid model, where only the solid region ($30 \times 30$ elements) is considered and its perimeter is directly treated as the open boundary to the external environment.
The second is a void-surrounded model, where the same solid region is embedded by a void region ($\phi=0$) discretized into $60 \times 60$ elements.
The calculated average surface temperatures for both configurations are summarized in \tref{tabVerificationVoid}.
The numerical results are in agreement with the analytical solution and confirm that the proposed discrete model conserves the total radiative energy emitted in all directions and that void regions transmit radiative heat flux without introducing artificial attenuation.
\begin{table}[htbp]
  \caption{Comparison of calculated dimensionless temperatures for the void verification models.}
  \label{tabVerificationVoid}
  \centering
  \begin{tabular}{lc}
    \hline
    Configuration                                         & Dimensionless temperature $T^*$ \\
    \hline
    Analytical solution (\eref{eqtempana})                & 7.07107                         \\
    Bare solid model (\fref{figVerificationVoid}(a))      & 7.07100                         \\
    Void-surrounded model (\fref{figVerificationVoid}(b)) & 7.07100                         \\
    \hline
  \end{tabular}
\end{table}
\par
\begin{figure}[htbp]
  \centering
  \begin{tikzpicture}[scale=1.1,color=black]
    \begin{scope}[shift={(0,0)}]
      \draw[thick, fill=red!20] (0,0) rectangle (2,2);
      \node[align=center] at (1, 1) {Solid $(\phi=1)$\\($30 \times 30$)};
      \foreach \x in {0.3, 1, 1.7} {
          \draw[->, red!80, thick, decorate, decoration={snake, amplitude=1.0, segment length=5, post length=1mm}] (\x, 2.3) -- (\x, 3.0);
        }
      \node[below, align=center, font=\bfseries] at (1, -0.5) {(a) Base solid};
    \end{scope}
    \begin{scope}[shift={(4.5,0)}]
      \draw[thick, fill=black!20] (-0.2,-0.2) rectangle (2.2,2.2);
      \draw[thick, fill=red!20] (0,0) rectangle (2,2);
      \node[align=center] at (1, 1) {Solid $(\phi=1)$\\($30 \times 30$)};
      \foreach \x in {0.3, 1, 1.7} {
          \draw[->, red!80, thick, decorate, decoration={snake, amplitude=1.0, segment length=5, post length=1mm}] (\x, 2.3) -- (\x, 3.0);
        }
      \draw[<-] (1.8, 2.07) -- (2.2,2.7) node[right, font=\small, align=left] {1 gray layer\\($\phi=0.5$)};
      \node[below, align=center, font=\bfseries] at (1, -0.5) {(b) Base solid + 1 gray layer};
    \end{scope}
    \begin{scope}[shift={(9.0,0)}]
      \draw[thick, fill=red!20] (-0.2,-0.2) rectangle (2.2,2.2);
      \draw[thick, dashed] (0,0) rectangle (2,2);
      \node[align=center] at (1, 1) {Solid $(\phi=1)$\\($30 \times 30$)};
      \foreach \x in {0.3, 1, 1.7} {
          \draw[->, red!80, thick, decorate, decoration={snake, amplitude=1.0, segment length=5, post length=1mm}] (\x, 2.3) -- (\x, 3.0);
        }
      \draw[<-] (1.8, 2.07) -- (2.2,2.7) node[right, font=\small, align=left] {1 solid layer\\($\phi=1$)};
      \node[below, align=center, font=\bfseries] at (1, -0.5) {(c) Base solid + 1 solid layer};
    \end{scope}
  \end{tikzpicture}
  \caption{Geometric settings for evaluating the effect of the intermediate density region.}
  \label{figVerificationGray}
\end{figure}
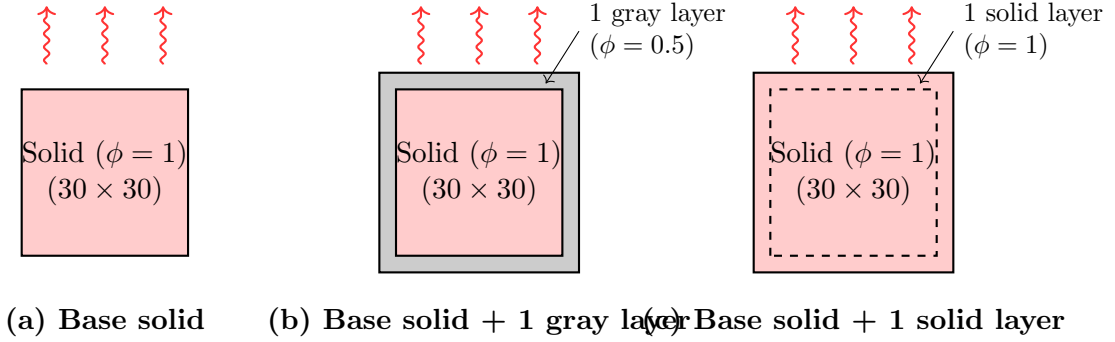
Next, the effect of intermediate density regions on radiative heat transfer is investigated.
To confirm that an intermediate-density region behaves as an intermediate state between solid and void with respect to radiative heat transfer, numerical analyses are conducted under three conditions, as illustrated in \fref{figVerificationGray}.
The base solid model (\fref{figVerificationGray}(a)) is constructed by a $30 \times 30$ solid elements located at the center of a computational domain discretized into $60 \times 60$ elements.
The base solid + 1 gray layer model (\fref{figVerificationGray}(b)) is constructed by adding a single-element layer of intermediate density ($\phi=0.5$) around the base solid model.
Similarly, the base solid + 1 solid layer model (\fref{figVerificationGray}(c)) is prepared by adding a single-element solid layer ($\phi=1.0$) around the base solid, which is equivalent to a $32 \times 32$ fully dense solid region.
The calculated steady-state temperatures are compared in \tref{tabVerificationGray}.
The gray layer model exhibits an equilibrium temperature between those of the base solid and the base solid + 1 solid layer models.
These results indicate that an intermediate-density layer behaves as an intermediate state between solid and void with respect to radiative heat transfer, suggesting that the proposed interpolation scheme provides a physically reasonable representation of the intermediate density region.
\begin{table}[htbp]
  \caption{Effect of the intermediate density region on the equilibrium dimensionless temperature.}
  \label{tabVerificationGray}
  \centering
  \begin{tabular}{lc}
    \hline
    Model                                                               & Dimensionless temperature $T^*$ \\
    \hline
    Base solid ($30\times 30$ elements) (\fref{figVerificationGray}(a)) & 7.07100                         \\
    Base solid + 1 gray layer           (\fref{figVerificationGray}(b)) & 6.98011                         \\
    Base solid + 1 solid layer          (\fref{figVerificationGray}(c)) & 6.95781                         \\
    \hline
  \end{tabular}
\end{table}

\subsection{Verification of mutual radiation and shadowing effects}\label{subsecBenchmark2}
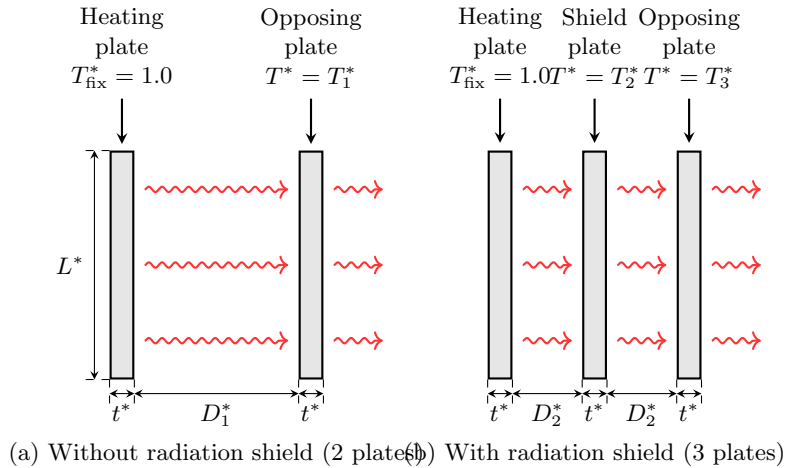
\begin{figure}[htbp]
  \centering
  \begin{tikzpicture}[
      font=\footnotesize,
      plate/.style={draw, thick, fill=gray!20, minimum width=3mm, minimum height=30mm, inner sep=0pt},
      dim/.style={<->, >=stealth, thin},
      pointer/.style={->, >=stealth, thick, shorten >=1mm},color=black
    ]
    \begin{scope}[xshift=0cm]
      \node[plate] (p1_a) at (0, 0) {};
      \node[plate] (p2_a) at (2.5, 0) {};
      \draw[thin, dashed] (p1_a.south west) -- +(0, -0.4);
      \draw[thin, dashed] (p1_a.south east) -- +(0, -0.4);
      \draw[thin, dashed] (p2_a.south west) -- +(0, -0.4);
      \draw[thin, dashed] (p2_a.south east) -- +(0, -0.4);
      \draw[thin, dashed] (p1_a.north west) -- +(-0.4, 0);
      \draw[thin, dashed] (p1_a.south west) -- +(-0.4,0);
      \draw[dim] ([xshift=-2mm]p1_a.south west) -- ([xshift=-2mm]p1_a.north west) node[midway, left] {$L^{*}$};
      \draw[dim] (p1_a.east |- 0,-1.7) -- (p2_a.west |- 0,-1.7) node[midway, below] {$D_1^{*}$};
      \draw[dim] (p1_a.south west |- 0,-1.7) -- (p1_a.south east |- 0,-1.7) node[midway, below] {$t^{*}$};
      \draw[dim] (p2_a.south west |- 0,-1.7) -- (p2_a.south east |- 0,-1.7) node[midway, below] {$t^{*}$};
      \draw[pointer] (0,2.2) node[above,align=center] {Heating\\plate\\ $T_{\rm fix}^*=1.0$} -- (p1_a.north |- 0,1.5);
      \draw[pointer] (2.5,2.2) node[above,align=center] {Opposing\\plate\\$T^*=T^*_1$} -- (p2_a.north |- 0,1.5);
      \foreach \y in {-1, 0, 1} {
          \draw[->, red!80, thick, decorate, decoration={snake, amplitude=1.0, segment length=5, post length=1mm}] (0.3,\y) -- ( 2.2, \y);
        }
      \foreach \y in {-1, 0, 1} {
          \draw[->, red!80, thick, decorate, decoration={snake, amplitude=1.0, segment length=5, post length=1mm}] (2.8,\y) -- ( 3.45, \y);
        }
      \node at (1.25, -2.5) {(a) Without radiation shield (2 plates)};
    \end{scope}
    \begin{scope}[xshift=5cm]
      \node[plate] (p1_b) at (0, 0) {};
      \node[plate] (p2_b) at (1.25, 0) {};
      \node[plate] (p3_b) at (2.5, 0) {};
      \draw[thin, dashed] (p1_b.south west) -- +(0, -0.4);
      \draw[thin, dashed] (p1_b.south east) -- +(0, -0.4);
      \draw[thin, dashed] (p2_b.south west) -- +(0, -0.4);
      \draw[thin, dashed] (p2_b.south east) -- +(0, -0.4);
      \draw[thin, dashed] (p3_b.south west) -- +(0, -0.4);
      \draw[thin, dashed] (p3_b.south east) -- +(0, -0.4);
      \draw[dim] (p1_b.east |- 0,-1.7) -- (p2_b.west |- 0,-1.7) node[midway, below] {$D_2^{*}$};
      \draw[dim] (p2_b.east |- 0,-1.7) -- (p3_b.west |- 0,-1.7) node[midway, below] {$D_2^{*}$};
      \draw[dim] (p1_b.south west |- 0,-1.7) -- (p1_b.south east |- 0,-1.7) node[midway, below] {$t^{*}$};
      \draw[dim] (p2_b.south west |- 0,-1.7) -- (p2_b.south east |- 0,-1.7) node[midway, below] {$t^{*}$};
      \draw[dim] (p3_b.south west |- 0,-1.7) -- (p3_b.south east |- 0,-1.7) node[midway, below] {$t^{*}$};
      \draw[pointer] (0, 2.2) node[above,align=center] {Heating\\plate\\$T_{\rm fix}^* =1.0$} -- (p1_b.north);
      \draw[pointer] (1.25, 2.2) node[above,align=center] {Shield\\plate\\$T^*=T^*_2$} -- (p2_b.north);
      \draw[pointer] (2.5, 2.2) node[above,align=center] {Opposing\\plate\\$T^*=T^*_3$} -- (p3_b.north);
      \foreach \y in {-1, 0, 1} {
          \draw[->, red!80, thick, decorate, decoration={snake, amplitude=1.0, segment length=5, post length=1mm}] (0.3,\y) -- ( 0.95, \y);
        }
      \foreach \y in {-1, 0, 1} {
          \draw[->, red!80, thick, decorate, decoration={snake, amplitude=1.0, segment length=5, post length=1mm}] (1.55,\y) -- ( 2.2, \y);
        }
      \foreach \y in {-1, 0, 1} {
          \draw[->, red!80, thick, decorate, decoration={snake, amplitude=1.0, segment length=5, post length=1mm}] (2.8,\y) -- ( 3.45, \y);
        }
      \node at (1.25, -2.5) {(b) With radiation shield (3 plates)};
    \end{scope}
  \end{tikzpicture}
  \caption{Computational models for the verification of mutual radiation and shadowing effect.}
  \label{figParallelPlates}
\end{figure}
Subsequently, the ability of the proposed model to evaluate mutual radiation between elements, as well as the shadowing effect caused by intervening structures, is examined.
As illustrated in \fref{figParallelPlates}, a system of parallel plates with a dimensionless length of $L^*$ and a dimensionless thickness of $t^*$ is considered.
All plate surfaces are allowed to radiate to the external environment, which is assumed to be at absolute zero ambient temperature ($T_{\text{env}}^* = 0$).
As illustrated in \fref{figParallelPlates}(a), the two-plate model consists of a fixed heating plate and an opposing plate separated by a dimensionless distance $D_1^*$.
The temperature of the heating plate is fixed at $T_{\rm fix}^*$, and the resulting equilibrium temperature of the opposing plate is denoted by $T_1^*$.
In the three-plate model shown in \fref{figParallelPlates}(b), to investigate the shadowing effect, a radiation shield is inserted midway between the heating plate and the opposing plate, each separated by $D_2^*=(D_1^*-t^*)/2$.
The equilibrium temperatures of the shield and the opposing plate are denoted by $T_2^*$ and $T_3^*$, respectively.
\par
The analytical solutions are derived based on thermodynamic energy balances and geometric view factors obtained via Hottel's crossed-string method \cite{hottel1967radiative, modest2021radiative}.
For two parallel plates of dimensionless length $L^*$ separated by a distance $D^*$, the view factor $F_{\rm face}(D^*)$ between the perfectly facing, zero-thickness surfaces is given by
\begin{align}
  F_{\rm face}(D^*)
  = \frac{2\sqrt{(D^*)^2 + (L^*)^2} - 2D^*}{2L^*}
  = \sqrt{\left(\frac{D^*}{L^*}\right)^2 + 1} - \frac{D^*}{L^*}.
\end{align}
Since the plates in the present benchmark possess a finite thickness $t^*$, radiation occurs from the entire perimeter.
Assuming uniform internal temperature, the total effective radiative surface area per unit out-of-plane depth is $A_{\rm tot}^* = 2L^* + 2t^*$.
Consequently, the overall view factor $F(D^*)$ can be calculated by scaling $F_{\rm face}(D^*)$ with the area ratio as follows:
\begin{align}
  F(D^*) = \frac{L^*}{2L^* + 2t^*} F_{\rm face}(D^*).
\end{align}
\par
For the two-plate model illustrated in \fref{figParallelPlates}(a), the heating plate is maintained at $T_{\rm fix}^*$, and the distance to the opposing plate is $D_1^*$, yielding the overall view factor $F_1 = F(D_1^*)$.
Under the assumption of a zero ambient temperature ($T_{\rm env}^* = 0$), the steady-state energy balance for the opposing plate dictates that the emitted energy equals the absorbed radiative energy:
\begin{align}
  A_{\rm tot}^*(T_1^*)^4 = F_1 A_{\rm tot}^* (T_{\rm fix}^*)^4 \quad \Rightarrow \quad T_1^* = T_{\rm fix}^* (F_1)^{1/4}.
\end{align}
\par
For the three-plate model illustrated in \fref{figParallelPlates}(b), a radiation shield at temperature $T_2^*$ is inserted midway between the heating plate and the opposing plate at temperature $T_3^*$.
The uniform distance between adjacent plates is $D_2^*$, yielding the overall view factor $F_2 = F(D_2^*)$.
The steady-state energy balances for the shield and the opposing plate are given by
\begin{align}
  \text{Shield plate:}   & \quad A_{\rm tot}^*(T_2^*)^4 = F_2 A_{\rm tot}^*(T_{\rm fix}^*)^4 + F_2 A_{\rm tot}^*(T_3^*)^4, \\
  \text{Opposing plate:} & \quad A_{\rm tot}^*(T_3^*)^4 = F_2 A_{\rm tot}^*(T_2^*)^4.
\end{align}
Solving this system of equations yields the equilibrium temperatures:
\begin{align}
  T_2^* & = T_{\rm fix}^* \left( \frac{F_2}{1 - F_2^2} \right)^{1/4},   \\
  T_3^* & = T_{\rm fix}^* \left( \frac{F_2^2}{1 - F_2^2} \right)^{1/4}.
\end{align}
By substituting the specific geometric parameters $L^* = 1.0$, $t^* = 0.01$, $D_1^* = 1.01$, and $D_2^* = 0.50$ and the fixed temperature ($T_{\rm fix}^* = 1.0$) into the above equations, the theoretical temperature values presented in \tref{tabVerificationParallel} are obtained.
\par
For this benchmark, the conduction-radiation parameter is set to $N_{\rm R} = 10^{-4}$.
The domain is discretized into elements of dimensionless size $0.01$, with $N_{\rm pt} = 10$ launch points per element face and $N_{\rm ang} = 100$ rays per launch point.
A comparison between the analytical solutions and the computational results obtained using the proposed radiative heat transfer analysis method is presented in \tref{tabVerificationParallel}.
In the configuration without the shield plate (\fref{figParallelPlates}(a)), the numerical temperature of the opposing plate ($T_1^*$) agrees with the analytical solution, confirming that both the mutual radiative exchange between the plates and the radiation to the external environment are accurately captured.
Moreover, when the radiation shield is inserted (\fref{figParallelPlates}(b)), the temperature of the opposing plate ($T_3^*$) is reduced in accordance with the analytical prediction due to the radiation shielding effect.
The maximum relative error observed between the analytical and numerical temperatures across all cases is approximately $0.5\%$.
These minor discrepancies are primarily attributed to the presence of a slight temperature gradient in the solid domain and the discretization error in the ray-tracing procedure.
Overall, these results indicate that the proposed framework can capture radiative heat transfer phenomena, including mutual irradiation, radiative escape to the external environment, and shadowing effect by intervening structures.
\begin{table}[htbp]
  \caption{Comparison of the analytical and numerical dimensionless temperatures for the parallel plate models.}
  \label{tabVerificationParallel}
  \centering
  \begin{tabular}{clcc}
    \hline
    Symbol    & Description                                                     & Analytical $T^*$ & Numerical $T^*$ \\
    \hline
    $T_{1}^*$ & Opposing plate (two-plate model: \fref{figParallelPlates}(a))   & 0.67174          & 0.67173         \\
    $T_{2}^*$ & Shield plate (three-plate model: \fref{figParallelPlates}(b))   & 0.76223          & 0.76627         \\
    $T_{3}^*$ & Opposing plate (three-plate model: \fref{figParallelPlates}(b)) & 0.56689          & 0.56690         \\
    \hline
  \end{tabular}
\end{table}

\subsection{Verification of conduction-radiation coupled heat transfer}\label{subsecBenchmark3}
\begin{figure}[htbp]
  \centering
  \begin{tikzpicture}[
      font=\footnotesize,
      dim/.style={<->, >=stealth, thin},
      pointer/.style={->, >=stealth, thick, shorten >=1mm},
      color=black
    ]
    \draw[thick, fill=gray!20] (0, 0) rectangle (7, 0.5);
    \node at (3.5, 0.25) {Solid};

    \draw[thin, dashed] (0,0) -- (0,-1.1);
    \draw[thin, dashed] (7,0) -- (7,-1.1);
    \node[below] at (0, -1.1) {$x^*=0$};
    \node[below] at (7, -1.1) {$x^*=1$};

    \draw[thin, dashed] (0, 0.5) -- (-0.8, 0.5);
    \draw[thin, dashed] (0, 0) -- (-0.8, 0);
    \draw[dim] (-0.5, 0) -- (-0.5, 0.5);
    \node[left] at (-0.5, 0.25) {$L_y^*$};

    \draw[pointer] (-0.6, 0.9) node[above] {$\bar{T}^* = 0$} -- (0, 0.25);

    \draw[pointer] (7.8, 0.25) node[right] {$\bar{q}^* = 1.0$} -- (7, 0.25);

    \foreach \x in {0.5, 2.0, 3.5, 5.0, 6.5} {
        \draw[->, red!80, thick, decorate, decoration={snake, amplitude=1.0, segment length=5, post length=1mm}] (\x, 0.55) -- (\x, 1.0);
        \draw[->, red!80, thick, decorate, decoration={snake, amplitude=1.0, segment length=5, post length=1mm}] (\x, -0.05) -- (\x, -0.5);
      }
    \node[above] at (3.5, 1.0) {Radiation to external environment ($T_{\text{env}}^* = 0$)};
    \node[below] at (3.5, -0.5) {Radiation to external environment ($T_{\text{env}}^* = 0$)};
  \end{tikzpicture}
  \caption{Computational model and boundary conditions for the one-dimensional radiating fin problem.}
  \label{figVerificationFinModel}
\end{figure}
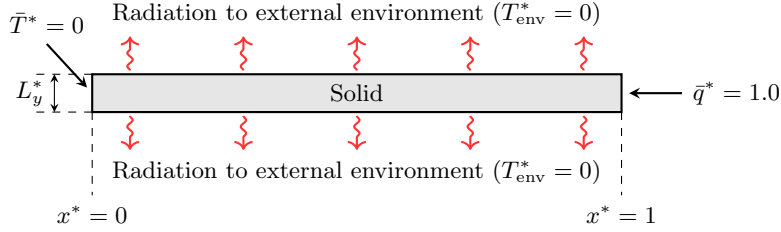
A one-dimensional radiating fin problem is considered to verify the capability of the proposed method in solving conduction-radiation coupled problems.
As illustrated in \fref{figVerificationFinModel}, the computational model consists of a rectangular domain extending from $x^* = 0$ to $x^* = 1$ with a small dimensionless thickness $L_y^* \ll 1$.
A Dirichlet boundary condition, $T^* = 0$, is applied at $x^* = 0$, while a constant inward dimensionless heat flux of $\bar{q}^{*}=1.0$ is applied at $x^* = 1$.
The top and bottom surfaces are exposed to the external environment with a dimensionless temperature of $T_{\rm env}^* = 0$, emitting heat via radiative heat transfer.
By balancing the one-dimensional heat conduction along the $x^*$-axis with the radiative heat emission from the top and bottom surfaces, the one-dimensional temperature field yields the following ordinary differential equation:
\begin{align}\label{eq1Dexample}
  \begin{aligned}
    L_y^*\frac{\text{d}^2 T^*(x^*)}{\text{d} x^{*2}} & = 2N_{\rm R}T^{*4}(x^*) &  & \text{in}\quad  x^*\in [0,1], \\
    T^*(x^*)                                         & =0                      &  & \text{at} \quad x^*=0   ,     \\
    \frac{\text{d}T^*(x^*)}{\text{d}x^*}             & =1.0                    &  & \text{at} \quad x^*=1.
  \end{aligned}
\end{align}
\par
\begin{figure}[htbp]
  \begin{center}
    \includegraphics[scale=1.0]{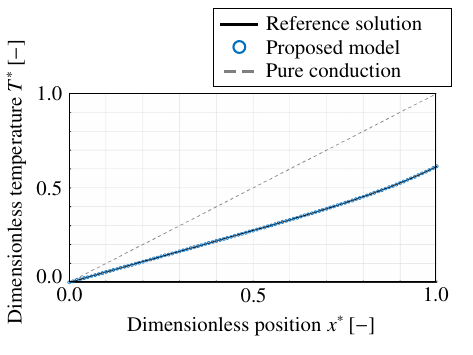}
  \end{center}
  \caption{Comparison of the dimensionless temperature along the $x^*$-axis for the one-dimensional conduction-radiation problem.}
  \label{figVerificationCoupling}
\end{figure}

For this benchmark, the conduction-radiation parameter $N_{\rm R}$ is set to $10^{-4}$.
The domain is discretized into $100 \times 1$ elements, with $N_{\rm pt} = 10$ launch points per element face and $N_{\rm ang} = 100$ rays per launch point.
\fref{figVerificationCoupling} compares the dimensionless temperature distributions along the $x^*$-axis obtained by the proposed method and the reference solution, the latter being obtained by solving \eref{eq1Dexample}.
To highlight the effect of conduction-radiation coupling, the theoretical temperature profile for pure heat conduction without radiation is also shown.
Compared with this linear profile, the actual temperature distribution bends downward due to radiative heat emission from the surfaces.
The dimensionless temperature predicted by the proposed method exhibits agreement with the reference solution (\eref{eq1Dexample}).
To quantitatively evaluate the accuracy, the relative error $e_{\rm rel}$ is calculated from
\begin{align}\label{eqerrorcheck}
  e_{\rm rel} = \frac{|T_{\rm proposed}^* - T_{\text{\eref{eq1Dexample}}}^*|}{T_{\text{\eref{eq1Dexample}}}^*},
\end{align}
where $T_{\rm proposed}^*$ and $T_{\text{\eref{eq1Dexample}}}^*$ are the dimensionless temperatures obtained from the proposed method and the reference solution, respectively.
The maximum relative error is $1.46 \times 10^{-4}$, confirming that the proposed method is capable of solving conduction-radiation coupled problems.

\section{Verification of the adjoint sensitivity}\label{appVerificationSensitivity}
\begin{figure}[htbp]
  \centering
  \begin{subfigure}[t]{0.38\linewidth} 
    \centering
    \includegraphics[scale=0.9]{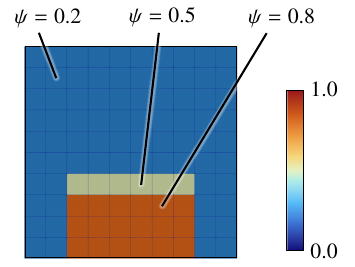}
    \caption{Prescribed design variable distribution}
    \label{subfigSens1}
  \end{subfigure}
  \hfill
  \begin{subfigure}[t]{0.6\linewidth}  
    \centering
    \includegraphics[scale=0.8]{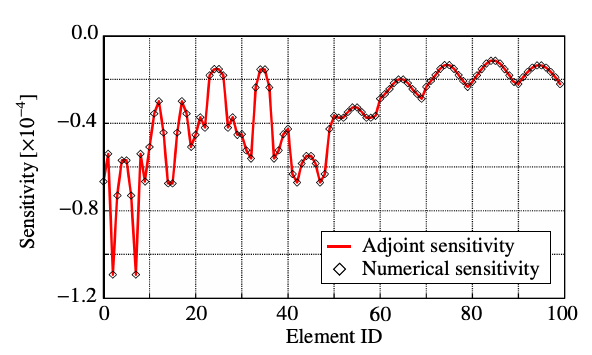}
    \caption{Comparison of the adjoint sensitivity and the numerical sensitivity} 
    \label{subfigSens3}
  \end{subfigure}
  \caption{Verification of the adjoint sensitivity.}
  \label{figSensitivity}
\end{figure}

The analytical sensitivity derived using the adjoint variable method is compared with the numerical sensitivity obtained by the finite difference method (FDM).
The two-dimensional problem described in \sref{subsecHeatSink2D} is adopted, with the conduction-radiation parameter set to $N_{\rm R}=1$.
To reduce the computational cost for the finite differentiation, the design domain $\Omega^*$ is coarsely discretized into $10 \times 10$ finite elements.
A prescribed density distribution containing intermediate-density regions is used for the verification, as shown in \fref{figSensitivity}(a).
The FDM sensitivity, denoted by $\left. \mathrm{d}J/\mathrm{d}\psi_{\df} \right|_{\rm FDM}$, is computed using a forward difference scheme with a perturbation of $\Delta\psi=10^{-7}$.
\par
As illustrated in \fref{figSensitivity}(b), the sensitivities obtained by the two methods show agreement across all elements.
For a quantitative evaluation, the relative error of the adjoint sensitivity and the numerical sensitivity, denoted as $e_{\rm rel}$, is evaluated as:
\begin{align}\label{eqsen_error}
  e_{\rm rel} =\left|  \frac{  \left. \frac{\text{d}J}{\text{d}\psi_{\df}} \right|_{\rm Adjoint} - \left. \frac{\text{d}J}{\text{d}\psi_{\df}} \right|_{\rm FDM} }{\left. \frac{\text{d}J}{\text{d}\psi_{\df}} \right|_{\rm FDM} }\right|.
\end{align}
In this verification, the maximum relative error ratio $e_{\rm rel}$ is approximately $5.82 \times 10^{-4}$, which is a sufficiently small value.
This result demonstrates that the adjoint sensitivity analysis of the proposed method is rigorously formulated.

\section{Effects of SIMP penalization on optimization results}\label{appInterpolation}
\newcommand{\SIMPscale}{1}
\begin{figure}[htbp]
  \centering
  \begin{subfigure}[t]{0.32\linewidth} 
    \centering
    \includegraphics[page=1,scale=\SIMPscale]{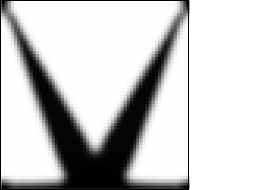}
    \caption{Optimized structure of $(p,q)=(1,1)$}
    \label{subfigSIMP11}
  \end{subfigure}
  \hfill
  \begin{subfigure}[t]{0.32\linewidth}  
    \centering
    \includegraphics[page=2,scale=\SIMPscale]{z511_SIMPCompare.pdf}
    \caption{Optimized structure of $(p,q)=(1,2)$}
    \label{subfigSIMP12}
  \end{subfigure}
  \hfill
  \begin{subfigure}[t]{0.32\linewidth}  
    \centering
    \includegraphics[page=3,scale=\SIMPscale]{z511_SIMPCompare.pdf}
    \caption{Optimized structure of $(p,q)=(1,3)$}
    \label{subfigSIMP13}
  \end{subfigure}\\\vspace{1em}
  \begin{subfigure}[t]{0.32\linewidth} 
    \centering
    \includegraphics[page=4,scale=\SIMPscale]{z511_SIMPCompare.pdf}
    \caption{Optimized structure of $(p,q)=(2,1)$}
    \label{subfigSIMP21}
  \end{subfigure}
  \hfill
  \begin{subfigure}[t]{0.32\linewidth}  
    \centering
    \includegraphics[page=5,scale=\SIMPscale]{z511_SIMPCompare.pdf}
    \caption{Optimized structure of $(p,q)=(2,2)$}
    \label{subfigSIMP22}
  \end{subfigure}
  \hfill
  \begin{subfigure}[t]{0.32\linewidth}  
    \centering
    \includegraphics[page=6,scale=\SIMPscale]{z511_SIMPCompare.pdf}
    \caption{Optimized structure of $(p,q)=(2,3)$}
    \label{subfigSIMP23}
  \end{subfigure}\\\vspace{1em}
  \begin{subfigure}[t]{0.32\linewidth} 
    \centering
    \includegraphics[page=7,scale=\SIMPscale]{z511_SIMPCompare.pdf}
    \caption{Optimized structure of $(p,q)=(3,1)$}
    \label{subfigSIMP31}
  \end{subfigure}
  \hfill
  \begin{subfigure}[t]{0.32\linewidth}  
    \centering
    \includegraphics[page=8,scale=\SIMPscale]{z511_SIMPCompare.pdf}
    \caption{Optimized structure of $(p,q)=(3,2)$}
    \label{subfigSIMP32}
  \end{subfigure}
  \hfill
  \begin{subfigure}[t]{0.32\linewidth}  
    \centering
    \includegraphics[page=9,scale=\SIMPscale]{z511_SIMPCompare.pdf}
    \caption{Optimized structure of $(p,q)=(3,3)$}
    \label{subfigSIMP33}
  \end{subfigure}
  \caption{Optimized structures obtained by varying the penalization parameters $p$ and $q$}
  \label{figSIMP}
\end{figure}

\newcommand{\SIMPAddscale}{1}
\begin{figure}[htbp]
  \centering
  \begin{subfigure}[t]{0.3\linewidth}
    \centering
    \includegraphics[scale=\SIMPAddscale,page=10]{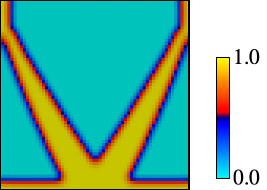}
    \caption{ Clear structure with $(p,q)=(2,2)$}
    \label{subfigSIMPadd1}
  \end{subfigure}
  \hspace{0.02\linewidth}
  \begin{subfigure}[t]{0.3\linewidth}
    \centering
    \includegraphics[scale=\SIMPAddscale,page=11]{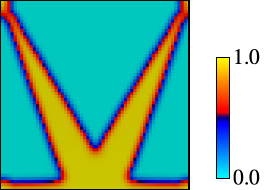}
    \caption{Branches connected by intermediate densities with $(p,q)=(1,1)$}
    \label{subfigSIMPadd2}
  \end{subfigure}
  \hspace{0.02\linewidth}
  \begin{subfigure}[t]{0.3\linewidth}
    \centering
    \includegraphics[scale=\SIMPAddscale,page=12]{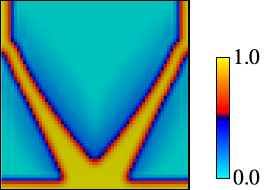}
    \caption{Diffuse intermediate densities with $(p,q)=(1,3)$}
    \label{subfigSIMPadd3}
  \end{subfigure}
  \caption{Density distributions visualized with a color map to highlight intermediate densities.}
  \label{figSIMPadd}
\end{figure}

\newcommand{\SIMPwidthB}{0.24\linewidth}
\newcommand{\SIMPscaleB}{0.85}
\begin{figure}[htbp]
  \centering
  \begin{subfigure}[t]{\linewidth}
    \centering
    \begin{minipage}[t]{\SIMPwidthB}
      \centering
      \includegraphics[page=1,scale=\SIMPscaleB]{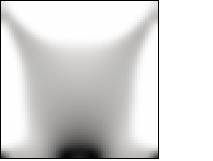}\\
      {\small Iter. 5}
    \end{minipage}\hfill
    \begin{minipage}[t]{\SIMPwidthB}
      \centering
      \includegraphics[page=2,scale=\SIMPscaleB]{z512_SIMPProcess.pdf}\\
      {\small Iter. 10}
    \end{minipage}\hfill
    \begin{minipage}[t]{\SIMPwidthB}
      \centering
      \includegraphics[page=3,scale=\SIMPscaleB]{z512_SIMPProcess.pdf}\\
      {\small Iter. 20}
    \end{minipage}\hfill
    \begin{minipage}[t]{\SIMPwidthB}
      \centering
      \includegraphics[page=4,scale=\SIMPscaleB]{z512_SIMPProcess.pdf}\\
      {\small Final}
    \end{minipage}
    \caption{Structural formation for $(p,q)=(2,1)$}
    \label{subfigProcess21}
  \end{subfigure}\vspace{1.5em}\\

  \begin{subfigure}[t]{\linewidth}
    \centering
    \begin{minipage}[t]{\SIMPwidthB}
      \centering
      \includegraphics[page=5,scale=\SIMPscaleB]{z512_SIMPProcess.pdf}\\
      {\small Iter. 5}
    \end{minipage}\hfill
    \begin{minipage}[t]{\SIMPwidthB}
      \centering
      \includegraphics[page=6,scale=\SIMPscaleB]{z512_SIMPProcess.pdf}\\
      {\small Iter. 10}
    \end{minipage}\hfill
    \begin{minipage}[t]{\SIMPwidthB}
      \centering
      \includegraphics[page=7,scale=\SIMPscaleB]{z512_SIMPProcess.pdf}\\
      {\small Iter. 20}
    \end{minipage}\hfill
    \begin{minipage}[t]{\SIMPwidthB}
      \centering
      \includegraphics[page=8,scale=\SIMPscaleB]{z512_SIMPProcess.pdf}\\
      {\small Final}
    \end{minipage}
    \caption{Structural formation for $(p,q)=(2,2)$}
    \label{subfigProcess22}
  \end{subfigure}\vspace{1.5em}\\

  \begin{subfigure}[t]{\linewidth}
    \centering
    \begin{minipage}[t]{\SIMPwidthB}
      \centering
      \includegraphics[page=9,scale=\SIMPscaleB]{z512_SIMPProcess.pdf}\\
      {\small Iter. 5}
    \end{minipage}\hfill
    \begin{minipage}[t]{\SIMPwidthB}
      \centering
      \includegraphics[page=10,scale=\SIMPscaleB]{z512_SIMPProcess.pdf}\\
      {\small Iter. 10}
    \end{minipage}\hfill
    \begin{minipage}[t]{\SIMPwidthB}
      \centering
      \includegraphics[page=11,scale=\SIMPscaleB]{z512_SIMPProcess.pdf}\\
      {\small Iter. 20}
    \end{minipage}\hfill
    \begin{minipage}[t]{\SIMPwidthB}
      \centering
      \includegraphics[page=12,scale=\SIMPscaleB]{z512_SIMPProcess.pdf}\\
      {\small Final}
    \end{minipage}
    \caption{Structural formation for $(p,q)=(2,3)$}
    \label{subfigProcess23}
  \end{subfigure}
  \caption{Structural formation for the selected interpolation parameters $(p,q)=(2,1)$, $(2,2)$, and $(2,3)$.}
  \label{figProcess}
\end{figure}
As discussed in \sref{secMaterialInterpolation}, the material penalization scheme affects the optimized structure.
This appendix examines the effects of the penalization parameter $p$ for the thermal conductivity $k^*(\phi)$ in \eref{eqInterpolationk} and the penalization parameter $q$ for the density interpolation function $f(\phi)$ associated with thermal radiation in \eref{eqInterpolationphi} on the resulting optimized structures.
The conduction-radiation parameter is set to $N_{\rm R}=1$, the volume constraint is set to $V_{\max}^*=0.3$, and a uniform initial density distribution of $\phi_{\te}=0.5$ is used.
Under these conditions, the optimization problem in \sref{subsecHeatSink2D} is performed for various combinations of the parameters $p$ and $q$.
\par
\fref{figSIMP} shows the optimized structures obtained for each parameter setting of $(p,q)$.
As shown in \frefs{subfigSIMP11} and \ref{subfigSIMP13}, using linear interpolation for the thermal conductivity ($p=1$) leads to intermediate-density regions. Specifically, the tips of the structural branches are connected by intermediate densities in \fref{subfigSIMP11}, whereas faint but widespread intermediate-density regions spread around the main structure in \fref{subfigSIMP13}.
For clarity, \fref{figSIMPadd} highlights the intermediate-density regions of these cases together with the case using $(p,q)=(2,2)$.
In contrast, introducing a penalty with $p \ge 2$ suppresses such intermediate densities, demonstrating the necessity of penalizing the thermal conductivity interpolation to obtain clear structures. This necessity of penalization is consistent with the pure conduction problems \cite{alberto2004topology,gersborg2006topologythermal}.
\par
Regarding the penalization of thermal radiation, the parameter $q$ strongly affects the structural formation process at the early optimization process.
\fref{figProcess} shows the structural evolution for $(p,q)=(2,1)$, $(2,2)$, and $(2,3)$ at 5-th, 10-th, 20-th, and final design iterations.
When the penalization of thermal conduction is stronger than that of thermal radiation ($p>q$), the structure initially forms in the central region of the design domain, as shown in \fref{subfigProcess21}.
This is because the stronger conduction penalty promotes the formation of conductive paths in the central region, thereby increasing the temperature of the radiative surface.
In contrast, when the radiation penalty is stronger ($p<q$), the structure initially forms along the boundaries of the design domain, as shown in \fref{subfigProcess23}.
Because a larger value of $q$ makes intermediate-density regions optically thinner, high-density regions first appear near the boundaries to increase the optical thickness in various viewing directions.
This difference in the structural formation at the early optimization process persists in the final optimized structures.
\par
To quantitatively compare the optimized structures, each design is binarized by assigning the solid phase ($\phi_{\te}=1$) to the elements with the highest densities until the volume fraction reaches $V^*=0.3$, while assigning the void phase ($\phi_{\te}=0$) to the remaining elements.
The objective function is then re-evaluated for each binarized structure.
The resulting objective function values are summarized in \tref{tabBinarizedJ}.
Except for the case with $(p,q)=(1,3)$ which has a large intermediate density region, the relative errors of the objective values are within 0.2\% across all settings. This result indicates that the penalization parameters have little influence on the final performance after binarization.
Based on this observation, $(p,q)=(2,2)$ is used in all numerical examples because it suppresses intermediate densities ($p\ge 2$) and applies the same penalization order to conduction and radiation.
\begin{table}[htbp]
  \centering
  \caption{Objective function $J$ after binarization for each penalization parameter setting $(p,q)$. Each structure is binarized at the volume fraction $V^*=0.3$.}
  \label{tabBinarizedJ}
  \begin{tabular}{cccc}
    \hline
          & $q=1$                 & $q=2$                 & $q=3$                 \\
    \hline
    $p=1$ & $6.8812\times10^{-4}$ & $6.8839\times10^{-4}$ & $7.0706\times10^{-4}$ \\
    $p=2$ & $6.8821\times10^{-4}$ & $6.8868\times10^{-4}$ & $6.8856\times10^{-4}$ \\
    $p=3$ & $6.8944\times10^{-4}$ & $6.8907\times10^{-4}$ & $6.8892\times10^{-4}$ \\
    \hline
  \end{tabular}
\end{table}

\section{Angular discretization error}\label{appdiscretizationError}
\begin{figure}[htbp]
  \begin{center}
    \includegraphics[scale=1.0]{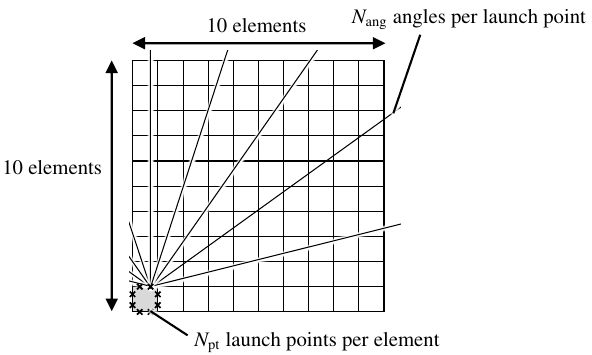}
  \end{center}
  \caption{Geometry for verifying the view factor calculation.}
  \label{figverificationViewFactor}
\end{figure}
This appendix assesses the angular discretization error in the proposed method. As shown in \fref{figverificationViewFactor}, a computational domain consisting of $10 \times 10$ elements is used for this verification. The view factor from the bottom-left element is calculated using the ray-tracing approach and compared with the analytical value obtained using Hottel's crossed-string method \cite{hottel1967radiative,modest2021radiative}. The discretization error is evaluated as the maximum relative error in the calculated view factors by varying the number of launch points per element face $N_{\rm pt}$ and the number of rays per launch point $N_{\rm ang}$.
\par
\tref{tabRayTracingConvergence} summarizes the maximum relative errors obtained for various combinations of $N_{\rm pt}$ and $N_{\rm ang}$.
As shown in the table, the error decreases as $N_{\rm pt}$ and $N_{\rm ang}$ increase, except for the coarsest angular resolution $N_{\rm ang}=10$, where the angular discretization error is dominant.
This monotonic convergence demonstrates that the proposed formulation can accurately evaluate the view factors when sufficiently fine ray-tracing discretizations are used.
\par
\begin{table}[htbp]
  \centering
  \caption{Maximum relative errors of the calculated view factors for different combinations of $N_{\rm pt}$ and $N_{\rm ang}$.}
  \label{tabRayTracingConvergence}
  \begin{tabular}{ccccc}
    \toprule
                 & \multicolumn{4}{c}{$N_{\rm ang}$}                                                                      \\
    \cmidrule{2-5}
    $N_{\rm pt}$ & 10                                & 100                  & 1000                 & 10000                \\
    \midrule
    1            & 1.836                             & 0.1874               & $8.249\times10^{-2}$ & $7.978\times10^{-2}$ \\
    10           & 1.836                             & $5.521\times10^{-2}$ & $7.901\times10^{-3}$ & $7.866\times10^{-4}$ \\
    100          & 1.836                             & $2.507\times10^{-2}$ & $2.631\times10^{-3}$ & $1.193\times10^{-4}$ \\
    1000         & 1.836                             & $1.851\times10^{-2}$ & $2.027\times10^{-4}$ & $7.037\times10^{-5}$ \\
    \bottomrule
  \end{tabular}
\end{table}
Furthermore, the impact of this angular discretization error on the optimization results is investigated.
\fref{figAngle} compares the optimized structures obtained using different angular discretization resolutions.
\fref{subfigAngle1}, which corresponds to the structure previously shown in \fref{subfigHeatSink2DNRm8}, is the optimized heat sink design obtained with $(p,q)=(2,2)$, $N_{\rm R}=10^{-8}$, $N_{\rm pt}=10$, and $N_{\rm ang}=100$.
As discussed in \sref{subsecHeatSink2D}, this structure exhibits irregular bending caused by angular discretization errors.
To verify that this irregularity is attributed to the discretization error, TO is performed for the same model using a finer discretization with $N_{\rm pt}=30$ and $N_{\rm ang}=300$.
\fref{subfigAngle2} shows the resulting optimized structure with the finer discretization.
It is observed that the irregular bending is eliminated, yielding a smoother structure.
This result demonstrates that a sufficiently fine discretization can mitigate the optimization instabilities that emerge under conditions with small $N_{\rm R}$.

\newcommand{\Anglescale}{1}
\begin{figure}[htbp]
  \centering
  \begin{subfigure}[t]{0.34\linewidth}
    \centering
    \includegraphics[scale=\Anglescale,page=1]{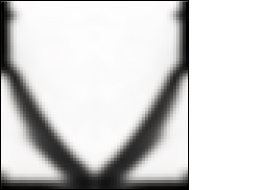}
    \caption{Structure with irregular bending obtained for $(p,q)=(2,2)$, $N_{\rm R}=10^{-8}$, $N_{\rm pt}=10$, \\and $N_{\rm ang}=100$}
    \label{subfigAngle1}
  \end{subfigure}
  \hspace{0.05\linewidth}
  \begin{subfigure}[t]{0.34\linewidth}
    \centering
    \includegraphics[scale=\Anglescale,page=2]{z341_AngleDiscretizationOpt.pdf}
    \caption{Straighter structure obtained for \\$(p,q)=(2,2)$, $N_{\rm R}=10^{-8}$, $N_{\rm pt}=30$, \\and $N_{\rm ang}=300$}
    \label{subfigAngle2}
  \end{subfigure}
  \caption{Comparison of the optimized structures obtained using different values of $N_{\rm pt}$ and $N_{\rm ang}$.}
  \label{figAngle}
\end{figure}

\bibliographystyle{elsarticle-num}
\bibliography{blackRadiation}

\end{document}